\documentstyle[12pt,preprint,aps,floats,axodraw]{revtex} 
\tighten 

\begin{document}

\setcounter{footnote}{1}

\preprint{ 
\noindent 
\begin{minipage}[t]{6in} 
\begin{flushright} 
 KEK-TH-701       \\
 SISSA 40/2000/EP \\
 UT-898           \\
hep-ph/0007018    \\ 
\vspace*{1.5cm} 
\end{flushright} 
\begin{center} 
\end{center} 
\end{minipage} 
} 
 
\draft 
 
\title{ Low-scale seesaw mechanisms for light neutrinos }
\author{
 Francesca Borzumati$^{\,a,b}$ and Yasunori Nomura$^{\,c}$}
\address{${}^{a}$  
 Theory Group, KEK, Tsukuba, Ibaraki 305-0801, Japan}
\address{${}^{b}$
 Scuola Internazionale Superiore di Studi Avanzati (SISSA),
 I--34013 Trieste, Italy}
\address{${}^{c}$ 
 Department of Physics, University of Tokyo, Tokyo 113-0033, Japan}

\maketitle 
 
\begin{abstract}

Alternatives to the seesaw mechanism are explored in supersymmetric
models with three right-handed or sterile neutrinos.  Tree-level Yukawa
couplings can be drastically suppressed in a natural way to give sub-eV
Dirac neutrino masses.  If, in addition, a $B\!-\!L$ gauge symmetry
broken at a large scale $M_G$ is introduced, a wider range of
possibilities opens up.  The value of the right-handed neutrino mass
$M_R$ can be easily disentangled from that of $M_G$.  Dirac and Majorana
neutrino masses at the eV scale can be generated radiatively through the
exchange of sneutrinos and neutralinos.  Dirac masses $m_D$ owe their
smallness to the pattern of light-heavy scales in the neutralino mass
matrix.  The smallness of the Majorana masses $m_L$ is linked to a
similar seesaw pattern in the sneutrino mass matrix.  Two distinct
scenarios emerge. In the first, with very small or vanishing $M_R$, the
physical neutrino eigenstates are, for each generation, either two light
Majorana states with a mixing angle ranging from very small to maximal,
depending on the ratio $m_D/M_R$, or one light Dirac state.  In the
second scenario, with a large value of $M_R$, the physical eigenstates
are two nearly unmixed Majorana states with masses $\sim m_L$ and 
$\sim M_R$. In both cases, the $(B\!-\!L)$-breaking scale $M_G$ is, in
general, much smaller than that in the traditional seesaw mechanism.

\end{abstract} 

\vfill  
 
\newpage  
\setlength{\parskip}{1.01ex}

\renewcommand{\thefootnote}{\arabic{footnote}}
\setcounter{footnote}{0}

\section{introduction} 
\label{intro} 
It has long been known that neutrinos are very light.  Whether they are
only light or completely massless, however, is a question which has
remained unanswered up until recently.

No direct measurement of neutrino masses exists as yet.  Nevertheless,
overwhelming evidence has been gathering in the past years of the fact
that neutrinos of different generations oscillate one into the
other~\cite{KAYSER}.  Oscillations have been observed by
atmospheric~\cite{SUPERK-ATM}, solar~\cite{SOLAR}, and
accelerator~\cite{LSND} neutrino experiments.  Quantum mechanics relates
oscillations among neutrinos of different flavor, $\nu_e$, $\nu_\mu$,
$\nu_\tau$, to nonvanishing values of mass for the neutrino mass
eigenstates, $\nu_1$, $\nu_2$, and $\nu_3$, which are linear
combinations of $\nu_e$, $\nu_\mu$, and $\nu_\tau$. The profusion of
experimental results on neutrino oscillations turns, therefore, into a
strong indication that neutrinos have mass.

This fact has profound consequences for particle physics, astrophysics,
and cosmology. Having mass, neutrinos may be a crucial ingredient of the
hot dark matter component that contributes to the mass density of the
Universe~\cite{DM-GALAXYFORM}. Moreover, if their mass does not exceed
the eV range, they may have played an important role in the formation of
the observed structure of galaxies~\cite{DM-GALAXYFORM}.  Far from being
a tested result, this value is a constraint to be kept into account by
model building, more stringent than the direct bounds from collider
experiments: $m_{\nu_e} \lesssim 5\,$eV, 
$m_{\nu_\mu} \lesssim 170\,$keV, and 
$m_{\nu_\tau} \lesssim 18\,$MeV~\cite{PDG}. It is also possible,
however, that some or all neutrinos are even lighter than a few eV.
Indeed, oscillation experiments give only an indication on the splitting
of neutrino masses squared, but leave the absolute size of neutrino
masses still open to speculation. They are not yet conclusive even on
the number of neutrino species undergoing oscillations.

The experimental situation can be schematically summarized as follows.
Atmospheric neutrino experiments from
Super-Kamiokande~\cite{SUPERK-ATM,SK-ATM-NU2000} indicate that muon
neutrinos $\nu_\mu$'s oscillate into different neutrinos, $\nu_x$'s. The
mass-squared splitting of the two mass eigenstate neutrinos (whose main
components are $\nu_\mu$ and $\nu_x$) is
$10^{-3}\,$eV$^2 < \Delta m^2 <10^{-2}\,$eV$^2$.  The neutrino $\nu_x$
seems to be predominantly a tau neutrino
$\nu_\tau$~\cite{SK-ATM-NU2000}. Results from the CHOOZ reactor
experiment~\cite{CHOOZ} strongly suggest that $\nu_x$ is unlikely to be
the electron neutrino $\nu_e$, although this possibility cannot be
excluded by the Super-Kamiokande Collaboration~\cite{SK-ATM-NU2000}.
Solar neutrino experiments~\cite{SOLAR} indicate that $\nu_e$'s
oscillate into some other type of neutrinos $\nu_{y}$'s.  If the MSW
mechanism~\cite{MSW} is used to explain the deficit in the solar
neutrino flux, the relevant mass-squared splitting is 
$ \Delta m^2 \sim 10^{-5}\,$eV$^2$ for the large and small mixing angle
solution, or $ \Delta m^2 \sim 10^{-7}\,$eV$^2$ for the LOW
solution~\cite{Gonzalez-Garcia:2001ae}.  It is 
$\Delta m^2 \sim 10^{-10}\,$eV$^2$ if oscillations in the vacuum between
Sun and Earth~\cite{VACUUM-OSC} are invoked.  Finally, the Liquid 
Scintillator Neutrino Detector (LSND) accelerator 
experiment~\cite{LSND} seems to support the hypothesis of a
$\nu_\mu \to \nu_e$ oscillation with corresponding mass-squared
splitting $\Delta m^2 \sim 1\,$eV$^2$.

Barring experimental systematic errors which may lead to inconsistent
results, these data cannot be explained by assuming the existence of
only three light neutrinos and that each of these oscillations is
described by a single $\Delta m^2$~\cite{KAYSER}. One simple way out is
to assume that there exist additional neutrinos, also light. Accurate
measurements of the invisible width of the $Z$ boson at the CERN $e+e-$
collider LEP have unequivocally shown that there are only three light
neutrinos coupling to the $Z$ boson~\cite{STERILE}. Other possible light
states, therefore, must have as main constituent neutrinos invisible for
the $Z$ boson, i.e., right-handed neutrinos.  When discussing neutrino
oscillations, these are often called sterile neutrinos, in contrast to
the left-handed components of SU(2) doublets, or active neutrinos.  If
sterile neutrinos exist, the results of solar neutrino experiments may
be viewed in terms of oscillations between electron and sterile
neutrinos. Up until recently, these were preferably accommodated in a
four-neutrino spectrum of the ``$2\!+\!2$'' variety. This spectrum
consists of two pairs of neutrinos separated by the LSND $\Delta m^2$, 
with the component of each pair
split, respectively, by the atmospheric and solar $\Delta m^2$. Recent
measurements of the solar neutrino flux by the Super-Kamiokande
Collaboration, however, seem to disfavor the possibility of $\nu_y$'s
being sterile neutrinos at the $95\%\,$C.L.~\cite{SK-SOL-NU2000}.  (A
different conclusion is reached in~\cite{GCPG}.) This fact, together
with a lower oscillation probability recently claimed by
LSND~\cite{NEWLSND}, resurrects the ``$3\!+\!1$'' neutrino
spectrum~\cite{THREEPLUSONE} previously excluded by the incompatibility
between LSND and negative searches for $\nu_e$ and $\nu_\mu$
disappearance. In this spectrum, one mainly-sterile neutrino is
separated by the LSND mass gap from the other three, which are mainly
active, roughly degenerate, and sufficient to explain atmospheric and
solar neutrino data. The ``$3\!+\!1$'' spectrum assumes a small mixing
between the active and sterile content of the four mass eigenstates. For
a model in which this spectrum can be easily implemented, see
Ref.~\cite{THREEPLUSONE_TH}.

The challenge, which particle physics is then faced with, is twofold.
On one side there is the longstanding issue of the smallness of neutrino
masses. Although clear experimental evidence that neutrinos are massive
is a recent acquisition, there has always been a strong theoretical
prejudice in this direction.  The exact vanishing of neutrino masses,
which is consistent with the standard-model (SM) particle content, would
nevertheless demand for a deeper explanation, possibly through a
symmetry principle, in an underlying theory of which the SM is an
effective limit.  Rejecting the possibility of vanishing neutrino masses
on theoretical ground, there has been a tremendous effort to explain why
these masses are much smaller than those of quarks and charged leptons.

Technically, two ways are known to obtain light neutrinos.  The first
follows the conventional mechanism for generating quark and
charged-lepton masses. The SM is enlarged to include three right-handed
neutrinos. These couple to the neutral components of the left-handed
leptonic doublets through Yukawa interaction terms.  After electroweak
symmetry breaking, there remain three light mass eigenstates $\nu_1$,
$\nu_2$, and $\nu_3$, which are Dirac fermions. For these neutrinos to
be at the eV scale, Yukawa couplings as small as $\sim 10^{-11}$ are
needed. The second known way of generating small neutrino masses is
through the seesaw mechanism~\cite{SEESAW1,SEESAW2}. This invokes the
existence of right-handed neutrinos with a very large mass 
$M_R \sim 10^{12}\,$GeV, whereas the couplings of the Yukawa neutrino
interactions are unsuppressed.  The smallness of the light neutrino
masses is then explained in terms of the seesaw pattern of light-heavy
scales in the neutrino mass matrix. The three light mass eigenstates
$\nu_1$, $\nu_2$, and $\nu_3$ are now Majorana fermions, with mass 
$\sim m_{\rm weak}^2/M_R$. The other three mass eigenstates, $n_1$,
$n_2$, and $n_3$, also of Majorana type, remain at the high scale $M_R$.
(On a different line of thought is the recent proposal of
Ref.~\cite{EXTRADIM}, where the smallness of neutrino masses is
explained by embedding the SM and right-handed neutrinos in a
higher-dimensional spacetime. In a scenario of Dirac neutrino masses,
for example, the Dirac Yukawa couplings between left-handed neutrinos
localized on the SM three-brane and the right-handed ones, residing in
the bulk, are suppressed by the volume factor of compactified
dimensions.)

The other challenge for theoretical particle physics has been
unexpectedly kindled by recent experimental results. The interpretation
of data coming from solar neutrino experiments in terms of oscillations
between active and sterile neutrinos requires extensions of the SM that
explain the existence of light mass eigenstates, other than $\nu_1$,
$\nu_2$, $\nu_3$, mainly corresponding to additional SU(2)$\times$ U(1)
singlets.  It is possible that the experimental situation will evolve in
the future in such a way that (i) the preliminary findings of the
Super-Kamiokande Collaboration on solar neutrino oscillations
disfavoring the sterile neutrino option are confirmed, (ii) the LSND
results are clearly excluded. The SNO experiment~\cite{SNO} will soon be
able to resolve the issue as to whether the electron neutrinos produced
in the Sun oscillate to active or sterile neutrinos.  In the meantime,
the Super-Kamiokande Collaboration will keep collecting more data on
solar neutrino oscillations that may possibly strengthen their
preliminary results.  Moreover, there is a number of already
operating~\cite{checkLSND-1} and planned~\cite{checkLSND-2} accelerator
experiments to seek for oscillations in the LSND region. If no signal is
found in these searches, the two $\Delta m^2$ of solar and atmospheric
neutrino oscillations can be easily accommodated in a scenario of three
light neutrinos. This, of course, does not mean that sterile neutrinos
do not exist. It may only mean that low-energy physics is blind to them,
either because they are heavy or because, even if light, they are very
weakly mixed with the active ones. Ideal detectors may turn out to be
astronomical ones, such as supernovas. Sterile neutrinos with a mass of
a few keV (or tens of keV) have been proposed to explain the motion of
pulsars born in supernova explosions~\cite{sternuPULSAR}. Light sterile
neutrinos, with mass above the eV scale, may also be of interest in
explaining supernova nucleosynthesis and big bang
nucleosynthesis~\cite{BBN-SNN}. If in the mass range of $0.1\,$keV, they
may also give rise to ``cool'' dark matter, as recently discussed
in~\cite{sternuDM}.

The natural setting for incorporating sterile neutrinos is that in which
six final Majorana states are seesaw generated.  Nevertheless, no
satisfactory theoretical model has been found so far to induce the
suppression factor $\sim m_{\rm weak}^2/M_R$ needed to have light
$\nu_1$, $\nu_2$, and $\nu_3$ states, while simultaneously keeping very
low the scale for some or all the right-handed neutrinos, with resulting
light states among $n_1$, $n_2$, and $n_3$.  (Infinitely many light
sterile neutrinos, however, are obtained in extra-dimensional mechanisms
for generating light neutrino masses~\cite{EXTRADIM}.)  In a typical
model, the seesaw right-handed neutrinos may participate in an
additional gauge interaction spontaneously broken at some scale
$M_G$. In the simplest known model, free from gauge anomalies and mixed
gauge and gravitational anomalies, this new interaction is $U(1)_{B-L}$
and the SM particle content is minimally enlarged by a neutral gauge
boson, three additional electroweak singlets, and a neutral
$(B\!-\!L)$-breaking Higgs boson. The mass of these new particles is
closely related to the spontaneous violation of ${B\!-\!L}$ and all
these particles are at the large scale $M_G$ ($\sim M_R$).

It is in general assumed that similar features hold also in the
supersymmetric version of such a model.  In this paper we explore other
possibilities that naturally present themselves in supersymmetric
models, in addition to the two outlined above, of generating small
neutrino masses.  Two mechanisms are described. In both, lowest-order
Yukawa operators are forbidden by imposing some horizontal symmetries.
Small neutrino masses can then be generated at the tree level, through
higher-dimensional nonrenormalizable operators, in the first mechanism,
or radiatively, through the virtual exchange of superpartners, in the
second one. The first mechanism, described in Sec.~\ref{mechanism-1},
does not require an additional $U(1)_{B-L}$ gauge interaction.  The
gauging of $U(1)_{B-L}$ is not needed either for the radiative
generation of Majorana masses for active neutrinos, but it is, on the
contrary, necessary to obtain radiative Dirac neutrino masses. For the
sake of a simpler and more coherent discussion, however, the radiative
mechanism is presented in the case of a gauged $U(1)_{B-L}$
symmetry. The modifications to the case without a $U(1)_{B-L}$ symmetry,
when applicable, are obvious and will be briefly commented upon.  The
radiative mechanism is introduced in Sec.~\ref{mechanism-2} and worked
out in Secs.~\ref{newseesaw},~\ref{seesawformulae}
and~\ref{effectiveinteractions}.  The patterns of physical neutrino
masses that can arise through both mechanisms are classified in
Sec.~\ref{neumasses}.  A possible embedding of these two mechanisms in
full-fledged models is outlined in Sec.~\ref{models}.  Up to this
section, it is implicitly or explicitly assumed that the signal of
supersymmetry breaking is transmitted from the hidden to the visible
sector through supergravity mediation. Comments will be made in
Sec.~\ref{models} on possible changes induced when this transmission is
realized through gauge mediation. The paper ends with some final
conclusions and remarks in Sec.~\ref{conclusion}.

\section{Supersymmetric Tree-Level Mechanism} 
\label{mechanism-1}
The class of models considered in this paper~\footnote{Tree-level
 mechanisms were also advocated in~\cite{BS} in the context of
 superstring theories, in~\cite{DN} in the context of supersymmetric
 models with gauge mediation of supersymmetry breaking, and in~\cite{TY}
 in the context of supergravity models.}
has three SU(2)$\times$ U(1) singlet superfields $\bar{N}$ in addition
to the three lepton doublets $L$, with neutral components $N$.  The
lowest-order operators giving rise to neutrino masses are
\begin{equation}
 W =  \frac{1}{4} \frac{h_\nu}{M_P} L H L H 
     + y_\nu \bar{N} L H
     + \frac{1}{2} M_R \bar{N} \bar{N}  \,,
\label{m_nu_spotential}
\end{equation}
with the scale $M_R$ naturally given by the Planck mass $M_P$. All these
operators can be forbidden by a horizontal discrete symmetry $Z_n$ in
conjunction with a continuous or discrete symmetry such as lepton
number. As for the first symmetry, it is sufficient to assign vanishing
$Z_n$ charges to the superfields $L$ and $H$ and $Z_n(\bar{N})=+1$ to
the superfields $\bar{N}$.  This assignment forbids the lowest-order
Yukawa operators, but leaves allowed higher-dimensional operators of
Yukawa type,
\begin{equation}
 W = \frac{k_\nu}{M_P} \, Z \bar{N} L H\,,
\label{softA}
\end{equation}
where $Z$ is a spurion superfield with charge $Z_n(Z) = -1$. The second
symmetry (different from $Z_2$) assigns the following charges to the
lepton superfields:
\begin{eqnarray}
  L(+1), \quad
  \bar{N}(-1), \quad
  \bar{E}(-1), \quad
\label{lepton-num}
\end{eqnarray}
and zero charge to all the other superfields. Therefore, it forbids both
Majorana mass operators in Eq.~(\ref{m_nu_spotential}) as well as the
higher-dimensional operators:
\begin{equation}
 W = \frac{1}{4} \frac{z_\nu}{M_P} \, Z Z \bar{N} \bar{N}\,.
\label{higher_majorana}
\end{equation}

In this scenario, the emerging picture is that of three very light
neutrino states of Dirac type.  The spurion field $Z$ acquires, in
general, a supersymmetry-conserving vacuum expectation value (VEV)
${\cal A}_Z$ and a supersymmetry-violating one ${\cal F}_Z$.  Through
operators analogous to that in Eq.~(\ref{softA}), the VEV ${\cal A}_Z$
gives contributions to Yukawa couplings for quarks and charged leptons
that are negligibly small, unless ${\cal A}_Z$ is at some large scale
$\mathop{}_{\textstyle \sim}^{\textstyle >} 10^{13}\,$GeV.  Acceptable
Yukawa couplings may, however, be generated for neutrinos.  Indeed, in a
class of O'Raifeartaigh-type models discussed in Sec.~\ref{models},
${\cal A}_Z$ is vanishing in the supersymmetric limit, but is shifted to
a value
\begin{equation}
{\cal A}_Z \simeq \frac{16\pi^2}{\lambda^3}  m_{\rm weak}
\label{az}
\end{equation}
by supersymmetry-breaking effects, where $\lambda$ is a dimensionless
coupling.  The induced Yukawa couplings
\begin{equation}
 y_\nu = 16\pi^2 \, \frac{k_\nu}{\lambda^3}
                 \frac{m_{\rm weak}}{M_P}
\label{Yukawa_Z}
\end{equation}
give rise to Dirac neutrino masses of order $10^{-2}\,$eV if $\lambda $
and $k_\nu$ are of ${\cal O}(1)$,~\footnote{We thank T.~Yanagida for
 pointing this out to us.}
i.e., in the range of the $\Delta m^2$ required by the MSW solution to
the solar neutrino deficit. Smaller values of $\lambda$ allow also
neutrino masses in the range of atmospheric neutrino experiments.  No
room is left for sterile neutrinos and the observed oscillation patterns
can be explained through flavor oscillations. This tree-level mechanism
for generating small neutrino Yukawa couplings is rather generic, in the
sense that it can be easily incorporated in any supersymmetric extension
of the SM model.

It is also possible to forbid the renormalizable operators in
Eq.~(\ref{m_nu_spotential}) leaving, however, allowed both operators in
Eqs.~(\ref{softA}) and~(\ref{higher_majorana}), as well as the
nonrenormalizable operator in~(\ref{m_nu_spotential}). One way to
achieve this is to couple the above $Z_n$ symmetry with an $R$-parity
symmetry under which $\bar{N}$ is odd and $Z$ is even. Dangerous terms
of type $M_P Z \bar{N}$ are then forbidden, while Majorana mass terms
$\sim M_P \bar{N} \bar{N}$ are avoided by requiring that $Z_n$ is not
$Z_2$. In this case, both Dirac and Majorana masses become possible at
the tree level: $\sim {\cal A}_Z v/M_P$ is obtained for Dirac masses;
$\sim {\cal A}_Z^2/M_P$ and $\sim v^2/M_P$, for Majorana masses for
right- and left-handed neutrinos, respectively, where $v$ is the VEV of
the neutral component of $H$.  The physical neutrino states are six, of
Majorana type: the three $n_i$ states have masses 
$\sim {\cal A}_Z^2 /M_P$, the three states $\nu_i$ are lighter, i.e., of
order of the Majorana mass for left-handed neutrinos, $\sim v^2/M_P$.
In the class of models in which supersymmetry breaking induces the value
of ${\cal A}_Z$ in Eq.~(\ref{az}), these masses range from the keV to
the sub-eV region, depending on the value of $\lambda$ and of the other
dimensionless couplings $z_\nu$, $h_\nu$, and $k_\nu$.  Thus, light
sterile neutrinos can be easily accommodated in such a scenario.

It should be noticed that the operator~(\ref{higher_majorana}) gives
rise to bilinear soft supersymmetry-breaking terms with large massive
couplings, $B M_R \sim {\cal A}_Z {\cal F}_Z /M_P$.  Such terms may
induce VEV's for the scalar components of the superfields ${\bar{N}}$ if
${\cal F}_Z$ is of the same order of the largest supersymmetry-breaking
VEV $\sim m_{\rm weak} M_P$. A value of ${\cal F}_Z$ low enough to avoid
the spontaneous breaking of lepton number decreases also the value of
the trilinear parameters $A \sim {\cal F}_Z /M_P$ generated by the
operator~(\ref{softA}). Nevertheless, large radiative contributions to
the Majorana mass for active neutrinos arise. We discuss these
contributions explicitly in the case of a gauged $U(1)_{B-L}$, when also
potentially large contributions to Dirac masses are induced.

\section{Supersymmetric Radiative Mechanism}
\label{mechanism-2}
If the gauge sector is enlarged as in a typical model that leads to the
seesaw mechanism, the phenomenology of the neutrino sector becomes much
richer.  The phenomenology of other sectors becomes also more
interesting.  The presence of the heavy scale $M_G$, typical of seesaw
models, for example, is felt also in other sectors. Some neutralinos
become very heavy and, depending on the details of the specific
realization of the model, also some supersymmetric partners of
neutrinos, or sneutrinos, can be at the same large scale of the
$B\!-\!L$ violation.

A suppression factor for small neutrino masses, therefore, does not
necessarily need to be induced by the right-handed neutrinos. Indeed,
there is at least one and, possibly, two mass matrices (neutralino and
sneutrino mass matrices) with the same seesaw pattern of light-heavy
scales that was only present in the neutrino mass matrix in the
nonsupersymmetric case. It is clear that the transmission to the
neutrino sector of a suppression factor induced in the neutralino and/or
sneutrino sector, may only happen at the quantum
level.~\footnote{Neutrino masses can be obtained radiatively also in
 other models. Among them, the Zee model and its supersymmetric version
 have long been known.  For recent discussions on this possibility and
 references to other radiative mechanism scenarios, see
 Ref.~\cite{ZEEmod}. See also Ref.~\cite{REFEREE} where Dirac masses are
 obtained at the quantum level in nonsupersymmetric models with extended
 gauge groups. Very popular are also $R_p$-violating models in which two
 neutrino masses are generated radiatively~\cite{RPVmodels}.} 
The advantage of such a procedure is obvious: the mechanism of mass
generation for the light neutrinos can be disentangled from the scale of
the sterile neutrinos, which does not need to be anymore that of the
$B\!-\!L$ violation.

If a program of radiative generation of neutrino mass operators is
implemented, the Yukawa couplings between active and sterile neutrinos
become redundant. As in the generic supersymmetric scenario outlined
above, they can be altogether forbidden.  Higher-dimension operators of
Yukawa type [see Eq.~(\ref{softA})] remain in general allowed and
provide Dirac neutrino masses naturally of order $10^{-2}\,$eV. If
larger physical masses are aimed at--for example, in the eV range--then
the largest contribution to the Dirac neutrino mass can be of radiative
origin, if the scale of $B-L$ breaking is not too large. If this is not
the case, both tree-level and radiative mechanisms complement each
other, providing comparable contributions. The radiative contribution to
the Majorana mass for active neutrinos is, on the contrary, always
larger than the tree-level contribution arising in the second tree-level
scenario described in Sec.~\ref{mechanism-1}. The ingredients necessary
for generating radiatively neutrino mass operators can be supplied by
soft supersymmetry-breaking terms.  The chiral-flavor violation needed
for a Dirac mass operator can be provided by trilinear soft
terms~\cite{BFPT,NirNEW}. For a Majorana mass operator, the correct
SU(2) structure can also be induced by the same trilinear terms, whereas
the explicit lepton-number violation can come from bilinear soft terms
in the scalar components of the singlets $\bar{N}$. In both cases, the
correct number of $R$ charges for Dirac and Majorana mass operators, is
provided by neutralino mass terms.

The supersymmetric models that we consider here contain a $B\!-\!L$
symmetry broken at some large scale $M_G$ and three additional SM
singlets $\bar{N}$ with tree-level mass $M_R$.  Dirac mass operators
mixing active and sterile neutrinos as well as Majorana mass operators
for active neutrinos are generated at the quantum level via loops with
virtual exchange of neutralinos and sneutrinos. Explicit expressions for
the corresponding couplings $m_D $ and $m_L$ are given in
Sec.~\ref{seesawformulae} after a detailed description of the neutralino
and sneutrino mass matrices in Sec.~\ref{newseesaw}. For unsuppressed
soft parameters $A$ and $B$ ($A \sim B \sim m_{\rm weak}$), Dirac masses
$m_D$ are of order $\sim (m_{\rm weak}/8 \pi^2)(m_{\rm weak}/M_G)^2$,
whereas Majorana ones are $\sim (m_{\rm weak}/8 \pi^2)(m_{\rm
weak}/M_R)^3$ for $M_R \gg m_{\rm weak}$ and 
$\sim (B M_R/m_{\rm weak}/8 \pi^2)$ for $M_R \ll m_{\rm weak}$.  (The
qualitative behavior of the radiative contribution to Majorana masses
obtained in the second scenario of Sec.~\ref{mechanism-1} is discussed
in Sec.~\ref{seesawformulae}.)  Of the resulting six eigenvalues, three
are certainly light. The other three are light or heavy depending on the
value of the tree-level Majorana masses $M_R$ for the singlets
$\bar{N}$.

Thus, two classes of possible scenarios emerge, discriminated by the
value of the mass parameter $M_R$.  In the first one $M_R$ is large,
say, $M_R\sim M_G$, and the resulting three states $n_i$ are of 
${\cal O}(M_G)$. The states $\nu_i$ can be easily of ${\cal O}$(eV) or
less. The main contribution to the corresponding eigenvalues comes from
the Majorana masses $m_L$, which dominate over $m_D^2/M_R$. Therefore,
the heavy $U(1)_{B-L}$ scale can be as light as a few hundred TeV.
[Notice that, if the lack of a dynamical explanation for such a large
scale $M_R$ is tolerated, the same neutrino spectrum can be obtained
without introducing a $U(1)_{B-L}$ symmetry.]

In the second class, $M_R$ is very light and Dirac masses play a much
more relevant role. The $n_i$ states can be as light as, or lighter
than, the $\nu_i$'s, which are at the eV scale if $M_G$ is
$\sim 10^7\,$GeV.  They can be heavier, as, for example, up to the scale
required to explain the motion of pulsars.  They can also be degenerate
with the $\nu_i$ states in the particular case of the exact vanishing of
$M_R$: in this case, for each generation $i$, the two Majorana states
$\nu_i$ and $n_i$ are equivalent to one Dirac neutrino.  The key
ingredients for this second class of scenarios are the lightness of
$M_R$ and the seesaw suppression of $m_D$ originating in the neutralino
mass matrix. Ways to render $M_R$ much smaller than $M_G$ are discussed
in Sec.~\ref{newseesaw}.

From the phenomenological point of view, $U(1)_{B-L}$ scales as low as
those needed for these two classes of scenarios are either too heavy or
marginally suited to give clear collider signatures due to the
additional neutral gauge boson or, possibly, heavy $\bar{N}$
states. Drastic deviations from the phenomenology of the minimal
supersymmetric standard model may, however, be observed in the scalar
sector if $M_R \lesssim m_{\rm weak}$.  Phenomenological consequences as
well as cosmological implications of these two scenarios are postponed
to later work.  Finally, in Sec.~\ref{effectiveinteractions}, we recall
that in a scenario of radiative neutrino masses, also Yukawa couplings
are generated radiatively.  These are now {\it naturally} very small:
i.e., $\sim m_D/m_{\rm weak}$ for the dimensionless Dirac coupling and
$m_L/m_{\rm weak}^2$ for the dimensionful Majorana coupling.  For Dirac
and Majorana masses in the eV range, they are in general larger than the
coupling induced by Yukawa operators of higher dimensionality.

\section{Radiative Seesaw Mechanisms}
\label{newseesaw}
Six massive Majorana neutrino states are usually obtained in a
three-family scenario with three left-handed neutrinos $\nu_L$ and three
right-handed ones $\nu_R$ with mass operators
\begin{equation}
 - {\cal  L}_{\rm mass} =  - \frac{1}{2} \nu_L^T C \, m_L \,\nu_L   + 
  \bar{\nu}_R \, m_D \, \nu_L  - \frac{1}{2} \nu_R^T \,C \, M_R^\ast \,\nu_R
  + {\rm h.c.} \,, 
\label{generallagr}
\end{equation}
where $C$ is the charge conjugation matrix.  Here and in the following,
all family indices are omitted for simplicity.  In the conventional
seesaw mechanism, with $m_D \simeq m_{\rm weak} \ll M_R$, the six
Majorana states are split into three light ones, 
$\nu \simeq \nu_L + \nu_L^{\,c}$, with mass 
$m_{\nu} \simeq - m_D \,(1/M_R) \,m_D^T $ and three heavy ones, 
$ n \simeq \nu_R + \nu_R^{\,c}$, with mass $m_n \simeq M_R$.

In supersymmetric models with violation of $B\!-\!L$ at some large scale
$M_G$, both Dirac and Majorana mass operators can be generated
radiatively through neutralino-sneutrino loops.  Suitable suppression
factors, with the task of keeping $m_D$ and $m_L$ well below the
electroweak scale, can be induced by the seesaw pattern of high-low
scales in the neutralino and/or sneutrino mass matrix.

The SM fields are charged under the $B\!-\!L$ symmetry, which, for
simplicity, is assumed to be a $U(1)$ symmetry. Nevertheless, the
following discussion can be easily generalized to larger gauge
groups. Under $U(1)_{B-L}$, quark, lepton, and Higgs chiral superfields
are charged in such a way that all Yukawa interaction terms are
invariant under $U(1)_{B-L}$.  A charge assignment that leads to
vanishing gauge anomalies is, however, unique~\cite{U1CHARGES}. It is
given by
$$
   X_Q =+1,\ \, X_{\bar{U}} =+1,\ \, X_{\bar{D}}=-3,\ \,
   X_H =-2,\ \, X_{\bar{H}} =+2,
$$
\begin{equation}
   X_L =-3,\ \, X_{\bar{E}} =+1,\ \, X_{\bar{N}} =+5.
 \label{bmlcharges}
\end{equation}
Neutrino Yukawa terms such as those in Eq.~(\ref{m_nu_spotential}) can
be forbidden at the tree level by the discrete $Z_n$ symmetry discussed
in Sec.~\ref{mechanism-1}.  On the other hand, soft
supersymmetry-breaking trilinear terms involving sneutrino fields can be
induced by superpotential terms such as those in Eq.~(\ref{softA}) when
the spurion field $Z$, which is neutral under $U(1)_{B-L}$ and has $Z_n$
charge $Z_n(Z) = -1$, acquires the supersymmetry-violating VEV 
${\cal F}_Z$.  As already explained in Sec.~\ref{mechanism-1}, the same
operator gives rise to tree-level Yukawa couplings when $Z$ acquires the
supersymmetry-conserving VEV ${\cal A}_Z$. In order to avoid too large
neutrino masses, a hierarchy between ${\cal A}_Z$ and ${\cal F}_Z$ must
exist. The possibility of implementing such a hierarchy, which was
already assumed in Sec.~\ref{mechanism-1}, will be discussed in
Sec.~\ref{models}.  The tree-level contributions can then be naturally
suppressed to give neutrino Yukawa couplings as small as $\sim 10^{-13}$
and Dirac masses well below the eV range.

Majorana mass terms $(1/2) M_R \bar{N} \bar{N}$ are assumed to exist at
the tree level. It is easy to see how $M_R$ can be completely
disentangled from $M_G$.  Majorana mass terms, indeed, are induced by
higher-dimensional operators in the superpotential [(different from
those considered in the case without a $U(1)_{B-L}$ symmetry discussed
in Sec.~\ref{mechanism-1}] that are neutral under $Z_n$ and
$U(1)_{B-L}$. For a generic $Z_n$ symmetry, these operators are
\begin{equation}
 W = \frac{1}{4} \frac{1}{\left(M_P\right)^{m+1}}  \,
   \bar{\Phi}^m Z Z \bar{N}\bar{N} \,, 
\label{majmassuno}
\end{equation}
whereas, if the discrete symmetry $Z_n$ is $Z_2$, the lowest-order
operators are
\begin{equation}
 W = \frac{1}{2} \frac{1}{\left(M_P\right)^{m-1}} \,
   \bar{\Phi}^m \bar{N}\bar{N} \,.
\label{majmassdue}
\end{equation}
In both cases, the field $\bar{\Phi}$ is a SM singlet that breaks
$U(1)_{B-L}$ and gets a VEV of ${\cal O}(M_G)$.  It is charged under
$U(1)_{B-L}$ with charge $X_{\bar{\Phi}}$. The power $m$ in
Eqs.~(\ref{majmassuno}) and~(\ref{majmassdue}) depends on
$X_{\bar{\Phi}}$, once the charge assignment for $\bar{N}$ is fixed as
in Eq.~(\ref{bmlcharges}).  It is clear, then, that the Majorana mass
$M_R$ does not need to be of the same order of $M_G$, and it can get
values over a wide range of scales below $M_G$.  Indeed, if the discrete
symmetry that forbids tree-level Yukawa terms is $Z_2$, it may be of
${\cal O}(M_G)$ if $m=1$. If $m=2$, depending on the value of $M_G$, it
may be a few order of magnitude above the electroweak scale or at the eV
scale (see possible values of $M_G$ discussed in Sec.~\ref{neumasses}).
It can also be extraordinarily suppressed if a generic $Z_n$ symmetry is
assumed and if $\bar{\Phi}$ has the $U(1)_{B-L}$ charge $X_{\bar{\Phi}}=
-1$. For the charge assignment in Eq.~(\ref{bmlcharges}), it is, for
example, $M_R \sim M_G^{10} \,{\cal A}_Z^2/(M_P)^{11} $.  Finally, if we
impose the symmetry in Eq.~(\ref{lepton-num}), the higher dimensional
operators in Eqs.~(\ref{majmassuno}) and~(\ref{majmassdue}) are
forbidden and the exact vanishing of $M_R$ is 
guaranteed.~\footnote{Depending on $X_{\bar{\Phi}}$, it is also possible
 that no Majorana mass term for $\bar{N}$ is allowed due to a residual
 unbroken discrete subgroup of $U(1)_{B-L}$, even if no other symmetry
 as that in Eq.~(\ref{lepton-num}) is imposed.}

Notice how the presence of a gauged $U(1)_{B-L}$ symmetry prevents, in
general, unwanted vacua with non-vanishing lepton number. Indeed, in
contrast to the situation described in the second scenario of
Sec.~\ref{mechanism-1}, no dangerously large bilinear soft
supersymmetry-breaking terms $B M_R$ are induced by the same operators
that generate $M_R$.

Tree-level Majorana mass terms for left-handed neutrinos are strongly
suppressed.  They may arise from superpotential operators:
\begin{equation}
 W = \frac{1}{4} \frac{1}{\left(M_P\right)^{m+1}}  \,
   {\Phi}^m  L H L H  \,, 
\label{majmassleft}
\end{equation}
where $\Phi$ is another SM singlet needed in addition to $\bar{\Phi}$ to
cancel $U(1)_{B-L}$ anomalies. It has $U(1)_{B-L}$ charge opposite to
that of $\bar{\Phi}$. As in the case of the operators in
Eqs.~(\ref{majmassuno}) and~(\ref{majmassdue}), also these Majorana mass
terms can be forbidden by the symmetry in Eq.~(\ref{lepton-num}).

\subsection{Sneutrino mass matrix}
\label{sneutrinomm}
As long as $M_R$ is different from zero, a bilinear and lepton
number-violating soft supersymmetry-breaking term 
$(1/2) B M_R \tilde{\bar{N}} \tilde{\bar{N}}$ is always induced.  Since
the parameter $B$ is, in general, at the electroweak scale, this term
may be very large or very small, depending on the value of $M_R$.  The
complete sneutrino mass potential is given by
\begin{eqnarray}
 - {\cal L}_{\rm mass} \ = && 
 m^2_{\tilde{l}}  \, \tilde{N}^\ast       \tilde{N}    + 
 m^2_{\tilde{\nu}}\, \tilde{\bar{N^\ast}} \tilde{\bar{N}} +      
 M_R^2            \, \tilde{\bar{N^\ast}} \tilde{\bar{N}} + 
 \frac{1}{2} M_R   \left(
   B      \tilde{\bar{N}}      \tilde{\bar{N}} 
  +B^\ast \tilde{\bar{N^\ast}} \tilde{\bar{N^\ast}} 
                   \right)                     \nonumber \\
                     && + 
  \left( A      v  \tilde{N}      \tilde{\bar{N}} 
       + A^\ast v  \tilde{N^\ast} \tilde{\bar{N^\ast}} 
        \right) \,,
\label{scalarmass}
\end{eqnarray}
where $v$ is the VEV of the relevant Higgs doublet. When redefining the
sneutrino fields as $\tilde{\nu}_L \equiv \tilde{N}$ and 
$\tilde{\nu}_R \equiv \tilde{\bar{N^\ast}}$, the above expression for 
$-{\cal L}_{\rm mass}$ becomes
\begin{eqnarray}
 - {\cal L}_{\rm mass} \ = && 
 m^2_{\widetilde{l}}  \,  \tilde{\nu}_L^\ast    \tilde{\nu}_L + 
 m^2_{\widetilde{\nu}}\,  \tilde{\nu}_R^\ast \, \tilde{\nu}_R +      
 M_R^2                \,  \tilde{\nu}_R^\ast \, \tilde{\nu}_R + 
 \frac{1}{2} M_R   \left(
   B \,   \tilde{\nu}_R^\ast \, \tilde{\nu}_R^\ast 
  +B^\ast \tilde{\nu}_R         \tilde{\nu}_R 
                   \right)                     \nonumber \\
                     && + 
  \left( A      v \, \tilde{\nu}_R^\ast \, \tilde{\nu}_L 
       + A^\ast v \, \tilde{\nu}_L^\ast    \tilde{\nu}_R 
        \right) \,.
\label{scalarmass-conv}
\end{eqnarray}

A nonvanishing soft parameter $B$ in Eq.~(\ref{scalarmass-conv}) 
induces a splitting in the $CP$-even and $CP$-odd sneutrino states. 
By decomposing $\tilde{\nu}_L$ and $\tilde{\nu}_R$ as 
$\tilde{\nu}_L \equiv 
   (\tilde{\nu}_{L\,+} +i \tilde{\nu}_{L\,-})/\sqrt{2}$ and 
$\tilde{\nu}_R \equiv 
   (\tilde{\nu}_{R\,+} +i \tilde{\nu}_{R\,-})/\sqrt{2}$,  
the sneutrino mass potential can, then, be expressed as
\begin{equation}
- {\cal L}_{\rm mass} = \frac{1}{2} 
 \left( \begin{array}{cccc}
   \tilde{\nu}_{L\,+}  & \tilde{\nu}_{R\,+} &
   \tilde{\nu}_{L\,-}  & \tilde{\nu}_{R\,-}       
        \end{array}\right)
 {\cal M}^2_{\rm sneut}
 \left( \begin{array}{c}
  \tilde{\nu}_{L\,+}   \\
  \tilde{\nu}_{R\,+}   \\
  \tilde{\nu}_{L\,-}   \\
  \tilde{\nu}_{R\,-}       
        \end{array}\right)\,. 
\label{snmassterms_gen}
\end{equation}
The matrix ${\cal M}^2_{\rm sneut}$ is, in general, a rather complicated
matrix. For one generation neutrino fields, i.e.,  ignoring the 
flavor-changing structure of $A$, $B$, $m_{\,\tilde{l}}$, and
$m_{\tilde{\nu}}$, ${\cal M}^2_{\rm sneut}$ is the following 
$4 \times 4$ matrix:
\begin{equation}
\vspace*{0.5truecm}
 \left( \!\! \begin{array}{cc|cc}
   m^2_{\,\tilde{l}}              
& \frac{1}{2} \left(A \! +\!\!A^\ast \! \right) v     
&  0                                 
& \frac{i}{2} \left(A^\ast\! \! -\!\!A \right) v 
\\[1.1ex]
  \frac{1}{2} \left(A \!+ \!\!A^\ast \! \right) v     
& \left(\! M_R^2 \!+\! \frac{1}{2} \left(B\! +\!\!B^\ast\!\right)\! M_R 
                  \! + \! m^2_{\tilde{\nu}}\right) 
& \frac{i}{2} \left(A \! -\!\!A^\ast\! \right) v 
& \frac{i}{2} \left(B^\ast\!\! -\!\!B \right)\! M_R            
\\ 
& & & 
\\[-1.2ex] \hline 
& & & 
\\[-1.2ex]
   0 
& \frac{i}{2} \left(A\! -\!\! A^\ast\! \right) v 
& m^2_{\,\tilde{l}}                    
& \frac{1}{2} \left(A\! +\!\! A^\ast\! \right) v    
\\[1.1ex]
  \frac{i}{2} \left(A^\ast \!\!- \!\!A \right) v 
& \frac{i}{2}   \left(B^\ast\!\!- \!\!B \right) \! M_R   
& \frac{1}{2} \left(A \!+ \!\!A^\ast \! \right) v         
& \left(\! M_R^2 \!-\! \frac{1}{2} \left(B\! +\!\!B^\ast\!\right) \! M_R 
                   + m^2_{\tilde{\nu}}\right) 
\\
        \end{array}\!\!\right)\,.
\vspace*{0.3truecm}
\label{sneumassmatrix}
\end{equation}

This matrix simplifies drastically when $A$ and $B$ are made real using
phase rotations of the $N$ and $\bar{N}$ superfields: it reduces, in
this case, to the block diagonal matrix
\begin{equation}
 {\cal M}^2_{\rm sneut} =  
 \left( \begin{array}{cc}
   {\cal M}^2_+ &  0_2                                        \\
    0_2         & {\cal M}^2_-      
        \end{array}\right)\,;
\hspace*{1truecm} {\rm with} \hspace*{0.5truecm} 
{\cal M}^2_{\pm}  = 
 \left( \begin{array}{cc}
   m^2_{\,\tilde{l}}   &   A v                                \\
   A v                 & M_R^2 \pm  B M_R + m^2_{\tilde{\nu}} 
        \end{array}\right)\,. 
\label{snmass_mat_real}
\end{equation}
(The same decomposition of $ {\cal M}^2_{\rm sneut}$ is found in
Ref.~\cite{GH}; see also~\cite{DK}.)

For large $M_R$ ($\sim M_G$), the two matrices ${\cal M}^2_{\pm}$ have
the typical pattern of light-heavy scales present in the neutrino mass
matrix in the conventional seesaw mechanism. Of the four eigenstates,
$\tilde{\nu}_{+,\,1}$, $\tilde{\nu}_{+,\,2}$, $\tilde{\nu}_{-,\,1}$, and
$\tilde{\nu}_{-,\,2}$, two ($\tilde{\nu}_{\pm,\,1}$) are at the
electroweak scale and two ($\tilde{\nu}_{\pm,\,2}$) are heavy, i.e., at
the scale $M_R$.  The two states in the light pair and those in the
heavy pair have masses that are split by terms proportional to
$B$. These splitting terms are suppressed by powers of $(m_{\rm
weak}/M_R)$. It is, indeed
\begin{eqnarray}
 m_{\tilde{\nu}_{\pm,\,1}}^2 
 &=& m_{\,\tilde{l}}^2 \left\{
    1 - (ac)^2 \beta^2 \pm (ac)^2 b \beta^3 \ +\ {\cal O}(\beta^4)
           \right\}
\nonumber \\
 m_{\tilde{\nu}_{\pm,\,2}}^2 
 &=& M_R^2 \left\{
      1 \pm  b \beta +d^2 \beta^2  \ + \ {\cal O}(\beta^4)  \right\} \,,   
\end{eqnarray}
where the definitions
\begin{equation}
 \beta \equiv \frac{m_{\,\tilde{l}}}{M_R} \,,
\hspace*{0.5truecm} 
  a    \equiv \frac{A}{m_{\,\tilde{l}}} \,,
\hspace*{0.5truecm} 
  b    \equiv \frac{B}{m_{\,\tilde{l}}} \,,
\hspace*{0.5truecm} 
  c    \equiv \frac{v}{m_{\,\tilde{l}}} \,,
\hspace*{0.5truecm} 
  d    \equiv \frac{m_{\tilde{\nu}}}{m_{\,\tilde{l}}} \,,
\label{defs}
\end{equation}
were used. At the same level of approximation, i.e.,  up to terms of
${\cal O}(\beta^4)$, the two diagonalization matrices of 
${\cal M}^2_{\pm}$, $ U_{\pm}$, are given by
\begin{equation}
 \left( \begin{array}{c}
    \tilde{\nu}_{L\,\pm} \\ \tilde{\nu}_{R\,\pm} 
        \end{array}\right)
 =  U_{\pm} 
 \left( \begin{array}{c}
    \tilde{\nu}_{\pm,\,1} \\ \tilde{\nu}_{\pm,\,2} 
        \end{array}\right)
 \equiv 
 \left( \begin{array}{cc}
 1     & -a c\left(-\beta^2 \pm b \beta^3\right)    \\
 ac \left(-\beta^2 \pm b \beta^3\right) &  1 
        \end{array}\right)
 \left( \begin{array}{c}
    \tilde{\nu}_{\pm,\,1} \\ \tilde{\nu}_{\pm,\,2} 
        \end{array}\right)\,.
\end{equation}
A Majorana mass operator radiatively induced by sneutrino-neutralino
loops must have a coupling proportional to the mass splitting of the
light or heavy sneutrino pairs, since this is induced by the
lepton-violating $B$ parameter. Notice that the splitting of the two
light eigenvalues comes only at order $(m_{\rm weak}/M_R)^3$.  It is,
therefore, conceivable that a Majorana mass term can be sufficiently
suppressed by several powers of $(m_{\rm weak}/M_R)$.

For very tiny $M_R$ and small $(B M_R / m_{\rm weak}^2)$, the correct
expansion parameter needed to calculate eigenvalues and eigenstates of
${\cal M}^2_{\pm}$ is the inverse of $\beta$, which we indicate by
$\gamma$. Since eigenvalues and eigenvectors are more involved in this
case, we express them in terms of those obtained at the zeroth order in
$\gamma$. For $M_R = 0$, ${\cal M}^2_{+}$ and ${\cal M}^2_{-}$ coincide,
and the sneutrino mass potential has the well-known form
\begin{equation}
 - {\cal L}_{\rm mass} =
 \left( \begin{array}{cc}
    \tilde{\nu}_L^\ast & \tilde{\nu}_R^\ast
        \end{array}\right)
 \left( \begin{array}{cc}
   m^2_{\,\tilde{l}} &  A v                \\
   A v               & m^2_{\tilde{\nu}}
        \end{array}\right)
 \left( \begin{array}{c}
    \tilde{\nu}_L \\ \tilde{\nu}_R
        \end{array}\right)\,,
\label{scalarssmatrix}
\end{equation}
once the phase of the $A$ parameter is rotated away.  Then, the
eigenvalues are given by
\begin{equation}
 m_{\tilde{\nu}_{1,2}}^2 = 
  \frac{1}{2}\left\{
  m_{\,\tilde{l}}^2 +  m_{\,\tilde{\nu}}^2
 \mp
  \sqrt{\left(m_{\,\tilde{l}}^2 -  m_{\,\tilde{\nu}}^2\right)^2
         + 4 \left( Av\right)^2 } \right\},
\label{eigen_nu}
\end{equation}
and the diagonalization matrix takes the following form:
\begin{equation}
 U  = 
 \sqrt{\displaystyle{
 \frac{ m_{\tilde{\nu}}^2   -m_{\tilde{\nu}_1}^2}
      { m_{\tilde{\nu}_2}^2 -m_{\tilde{\nu}_1}^2}}}
 \left( \begin{array}{cc}
  1    &  
\displaystyle{\frac{Av}{ m_{\tilde{\nu}}^2 -m_{\tilde{\nu}_1}^2}} \\
 -
\displaystyle{\frac{Av}{ m_{\tilde{\nu}}^2 -m_{\tilde{\nu}_1}^2}} 
      & 1 
        \end{array}\right)\,.
\end{equation}

If $\gamma$ is nonzero, the masses for the $CP$-even and $CP$-odd
sneutrino states split, and the eigenvalues of ${\cal M}^2_{\pm}$ now
become
\begin{eqnarray}
 m_{\tilde{\nu}_{\pm,\,1}}^2 
 &=& m_{\tilde{\nu}_{1}}^2  \ \pm \
 \frac{ m_{\tilde{\nu}_2}^2 -m_{\tilde{\nu}}^2}
      { m_{\tilde{\nu}_2}^2 -m_{\tilde{\nu}_1}^2} m_{\tilde{l}}^2 \, b \gamma 
 + {\cal O}(m_{\tilde{l}}^2 \gamma^2)
 \,,
\nonumber \\[1.1ex]
 m_{\tilde{\nu}_{\pm,\,2}}^2 
 &=& m_{\tilde{\nu}_{2}}^2 \ \pm \
 \frac{ m_{\tilde{\nu}}^2 -m_{\tilde{\nu}_1}^2}
      { m_{\tilde{\nu}_2}^2 -m_{\tilde{\nu}_1}^2} m_{\tilde{l}}^2 \, b \gamma 
 + {\cal O}(m_{\tilde{l}}^2 \gamma^2)
 \,, 
\label{sneu_eigenvalues_two}
\end{eqnarray}
with $b$ defined in Eq.~(\ref{defs}). At the same level of
approximation, the corresponding diagonalization matrices $ U_{\pm}$ are
\begin{equation}
 U_{\pm}  = 
 \sqrt{\displaystyle{
 \frac{ m_{\tilde{\nu}}^2   -m_{\tilde{\nu}_1}^2}
      { m_{\tilde{\nu}_2}^2 -m_{\tilde{\nu}_1}^2}}}
 \left( \begin{array}{cc}
  1 + x_{\pm}    &  
\displaystyle{\frac{Av}{ m_{\tilde{\nu}}^2 -m_{\tilde{\nu}_1}^2}} - y_{\pm} \\
 -
\displaystyle{\frac{Av}{ m_{\tilde{\nu}}^2 -m_{\tilde{\nu}_1}^2}} + y_{\pm} 
      & 1 + x_{\pm} 
        \end{array}\right) \,,
\label{sneu_eigenvectors_two}
\end{equation}
where $x_{\pm}$ and $y_{\pm}$ are given by
\begin{equation}
 x_{\pm}  =  
 \pm \frac{(Av)^2} 
      { \left(m_{\tilde{\nu}}^2  \! -\!m_{\tilde{\nu}_1}^2 \right) \!\!
        \left(m_{\tilde{\nu}_2}^2\! -\!m_{\tilde{\nu}_1}^2 \right)^2}   
 \, m_{\tilde{l}}^2 \, b \gamma  
 \,,
\hspace*{0.7truecm}
 y_{\pm}  =  x_{\pm} 
   \left\{
 \frac{Av}{m_{\tilde{\nu}}^2 \! -\! m_{\tilde{\nu}_1}^2} + 
 \frac{m_{\tilde{\nu}}^2     \! -\! m_{\tilde{l}}^2}{Av}
    \right\}   \,.
\label{sneu_defs_two}
\end{equation}
Differently than in the case with opposite hierarchy, i.e., 
$M_R \gg m_{\rm weak}$, here the first nonvanishing splitting of the two
pairs of sneutrino eigenvalues is linear in the suppression parameter
$(M_R/m_{\rm weak})$.

In the second scenario discussed in Sec.~\ref{mechanism-1} radiative
contributions to the Majorana mass for left-handed neutrinos are
possible and potentially large. In this scenario, $M_R$ is small, i.e.,
$M_R \ll m_{\rm weak}$, but $B M_R$, induced by the
operator~(\ref{higher_majorana}), is larger than all other entries in
the sneutrino mass matrix if ${\cal F}_Z$ is of the order of the maximal
VEV inducing supersymmetry breaking.  For coefficients $z_\nu$ of 
${\cal O}(1)$, a value of ${\cal F}_Z$ smaller than the typical VEV
$\sim m_{\rm weak} M_P$ is required in order to avoid a tachyonic
sneutrino state. When ${\cal F}_Z$ is minimally reduced, the splittings
in the two pairs of eigenvalues of the two matrices ${\cal M}^2_{\pm}$
is of $\lesssim {\cal O}(m_{\rm weak}^2)$. However, it is easy to
convince oneself, at least {\it a posteriori}, that the requirement
$m_{\nu_i} \lesssim 1\,$eV implies $B M_R < m_{\rm weak}^2$. Therefore,
the approximated expressions in
Eqs.~(\ref{sneu_eigenvalues_two}),~(\ref{sneu_eigenvectors_two}),
and~(\ref{sneu_defs_two}) still hold in this case.

\subsection{Neutralino mass matrix}
\label{neutralinomm}
The neutralino sector of this model is also more involved than in the
minimal supersymmetric extension of the SM. The spontaneous breaking of
$U(1)_{B-L}$ is encapsulated in the superpotential terms
\begin{equation}
 W = \frac{1}{\sqrt{2}} \, Y
    \left( \Phi \bar{\Phi} - v_{\Phi}^2 \right) \,.
\label{superhiggs}
\end{equation}
Here, $Y$ is a chiral superfield neutral under $U(1)_{B-L}$ that forces
$\Phi$ and $\bar{\Phi}$ to acquire VEV's $\langle \Phi \rangle = \langle
\bar{\Phi} \rangle = v_{\Phi}$, which can be assumed real and positive.
The two fields $\Phi $ and $\bar{\Phi}$ can be reexpressed in terms of a
Goldstone chiral multiplet $\Psi$, which is absorbed by the $U(1)_{B-L}$
gauge multiplet $X$, and a chiral field $K$.  The effect of this
super-Higgs mechanism is that the gauge multiplet acquires mass
\begin{equation}
 M_G = 2 g_X |X_\Phi| v_{\Phi} \,,
\end{equation}
where $g_X$ is the $U(1)_{B-L}$ gauge coupling and $X_\Phi$ the
$U(1)_{B-L}$ charge of the field $\Phi$. The superpotential
in~(\ref{superhiggs}) reduces to
\begin{equation}
 W = \frac{1}{2\sqrt{2}} Y K^2 + v_{\Phi} Y K \,.
\label{brokenphase}
\end{equation}
(Notice that a potentially dangerous coupling $Y H \bar{H}$, which could
shift the scale of the SM breaking up to $M_G$, can be easily forbidden
by making use of an $R$ symmetry.)

The implementation of a spontaneous breaking of $U(1)_{B-L}$ requires
therefore the inclusion of at least three SM neutral singlet
superfields. As a consequence, new neutralino states are present in
this model in addition to the usual $\tilde{B}$, $\tilde{W}_3$,
$\tilde{H}_0$, and $\tilde{\bar{H}}_0$.  For the three new singlet
superfields $\Phi$, $\bar{\Phi}$, and $Y$, the new neutralino states
are $\tilde{\Psi}$, $ \tilde{K}$, $\tilde{Y}$, and the $U(1)_{B-L}$ 
gaugino $\tilde{X}$. On the basis 
$ \left(
 \tilde{B}   \, \tilde{X}   \, \tilde{\Psi} \vert 
 \tilde{K}   \, \tilde{Y}   \, \vert\vert  \,
 \tilde{W}_3 \, \tilde{H}_0 \, \tilde{\bar{H}}_0   \right)^T$,
the neutralino mass matrix, when normalized as 
\begin{equation}
- {\cal L}_{\rm mass} = 1/2 
 \left(
 \tilde{B}   \, \tilde{X}   \, \tilde{\Psi} \vert 
 \tilde{K}   \, \tilde{Y}   \, \vert\vert  \,
 \tilde{W}_3 \, \tilde{H}_0 \, \tilde{\bar{H}}_0   \right)
  {\cal M}_{\rm neutr}  
 \left(
 \tilde{B}   \, \tilde{X}   \, \tilde{\Psi} \vert 
 \tilde{K}   \, \tilde{Y}   \, \vert\vert  \,
 \tilde{W}_3 \, \tilde{H}_0 \, \tilde{\bar{H}}_0   \right)^T\,,
\end{equation}
acquires the form
\begin{equation}
  {\cal M}_{\rm neutr}  = 
  \left( \begin{array}{ccc|cc||ccc}
   m_{\widetilde{B}} & m_{\rm mix}       & 0   
 & 0                 & 0 
 & 0                 & gv                & gv 
\\[1.01ex]
   m_{\rm mix}       & m_{\widetilde{X}} & M_G 
 & 0                 & 0 
 & 0                 & gv                & gv 
\\[1.01ex]
   0                 & M_G               & 0   
 & 0                 & 0 
 & 0                 & 0                 & 0 
\\[1.01ex] \hline & & & & & & & 
\\[-0.99ex]
   0                 & 0                 & 0 
 & 0                 & v_{\Phi}
 & 0                 & 0                 & 0 
\\[1.01ex]
   0                 & 0                 & 0 
 & v_{\Phi}          & 0
 & 0                 & 0                 & 0 
\\[1.01ex] \hline\hline & & & & & & & 
\\[-0.99ex]
   0                 & 0                 & 0   
 & 0                 & 0
 & m_{\widetilde{W}} & gv                & gv 
\\[1.01ex] 
   gv                & gv                & 0   
 & 0                 & 0 
 & gv                & 0                 & \mu
\\[1.01ex]
   gv                & gv                & 0   
 & 0                 & 0 
 & gv                & \mu               & 0 
\\ 
\end{array} \right)\,. 
\label{neutral_mat_expl}
\end{equation}
In Eq.~(\ref{neutral_mat_expl}), the two-component notation for the
spinor fields is used.  The entry $M_G$ is the Dirac-type mass that
couples $\tilde{X}$ to $\tilde{\Psi}$, whereas the Majorana-type mass
$m_{\widetilde{X}}$ for the gaugino $\tilde{X}$ is a soft
supersymmetry-breaking parameter of the order of the weak scale. The
entry $m_{\rm mix}$, also at the electroweak scale, is a mass terms that
mixes the U(1)$_Y$ and U(1)$_{B-L}$ gauginos.  The generic symbol $g v$
stays here for the product of a SU(2)$\times$ U(1) gauge coupling times
one of the two SM VEV's and some normalization factor, and it is not
always the same in the different entries of ${\cal M}_{\rm neutr}$.  The
diagonalization of this matrix through a unitary transformation
($D^\dagger {\cal M}_{\rm neutr} D$) leads to eight mass eigenstates
$\tilde{\chi}^0_i$, related to the current eigenstates through
\begin{equation}
 \left( \begin{array}{ccc|cc||ccc}
 \tilde{B}   &  \tilde{X}   &  \tilde{\Psi} & 
 \tilde{K}   &  \tilde{Y}   & 
 \tilde{W}_3 &  \tilde{H}_0 &  \tilde{\bar{H}}_0 
 \end{array} \right)^T 
= D
 \left( \begin{array}{cccccccc}
   \tilde{\chi}^0_1  &  \tilde{\chi}^0_2 &
   \tilde{\chi}^0_3  &  \tilde{\chi}^0_4 &
   \tilde{\chi}^0_5  &  \tilde{\chi}^0_6 &
   \tilde{\chi}^0_7  &  \tilde{\chi}^0_8 
 \end{array} \right)^T \,.
\label{neutreigenstates}
\end{equation}
Of these mass eigenstates, two are degenerate with mass $\pm v_{\Phi}$,
and two are almost degenerate, at the scale $M_G$. Their masses differ
by an overall sign; the splitting in the absolute value of these masses
is of order $m_{\rm weak}$. The remaining four eigenstates are at the
electroweak scale.

In order to visualize the relevance of these states for the radiative
generation of neutrino masses, in particular of Dirac type, it is
convenient to take the limit $v \to 0$ in the neutralino mass matrix.
Mixing terms due to SU(2) Higgsino-gaugino couplings as well as
SU(2)-gaugino-U(1)-gaugino couplings, which are of order
$(m_Z/\tilde{m})^2$, with $\tilde{m}$ a soft supersymmetry breaking
parameter, are in this case neglected. In this limit, ${\cal M}_{\rm
neutr}$ reduces to a block-diagonal matrix. The lower $3 \times 3$ block
has no relevance for the radiative generation of neutrino masses, since
no $SU(2)$ Higgsino-neutrino-sneutrino couplings are possible in the
absence of tree-level neutrino Yukawa couplings. The central $2\times 2$
block turns out to give a vanishing contribution to neutrino
masses. Therefore, in this limit, the $3 \times 3$ upper block is the
only one of interest to induce Dirac neutrino mass operators.  Fermionic
and scalar components of the singlets $\bar{N}$, indeed, couple only to
the gaugino $\tilde{X}$ and the field $\tilde{K}$. However, once the
rotation~(\ref{neutreigenstates}) is performed, all mass eigenstates
$\tilde{\chi}^0_i$ participate in the vertex 
$\bar{\nu}_R \tilde{\chi}^0_i \tilde{\nu}_R$, with couplings that are
larger for the heavy $\tilde{\chi}^0_i$ states. Two of these
$\tilde{\chi}^0_i$ states are a linear combination of $\tilde{K}$ and
$\tilde{Y}$ only, obtained through a rotation at 45$^0$ of $\tilde{K}$
and $\tilde{Y}$. These states have exactly opposite masses 
$\pm v_{\Phi}$ and their contributions to Dirac mass operators cancel
identically.  A similar feature holds for the contribution coming from
the two states that are mainly $\tilde{X}$ and $\tilde{\Psi}$. In this
case, however, the cancellation is not exact, but it is spoiled by
powers of the suppression factor $(m_{\rm weak}/M_G)$, where 
$m_{\rm weak}$ can be any of the remaining parameters in the upper
$3\times 3$ block of ${\cal M}_{\rm neutr} $, i.e. $m_{\widetilde{B}}$,
$m_{\rm mix}$, or $m_{\widetilde{X}}$. It turns out that the lowest
order suppression factor comes with power {\it 2}. As will be stressed
later, this fact has important phenomenological consequences.

We close this section by giving explicitly eigenvalues and eigenvectors
of the $3 \times 3$ upper block in the neutralino mass
matrix~(\ref{neutral_mat_expl}), assuming that its elements are
real. The three eigenvalues are
\begin{equation}
 m_{\tilde{\chi}_0^0}   =  m_{\widetilde{B}}\,,    
\hspace*{1.0truecm}
 m_{\tilde{\chi}_\pm^0} =  M_G 
     \left\{ \pm 1 + \alpha \frac{x}{2} 
                \pm \alpha^2 \left(\frac{x^2}{8} + \frac{z^2}{2} \right) 
     \right\} \,,                
\label{eigenvalues}
\end{equation}
where $\alpha$, $z$, and $x$ are
\begin{equation}
 \alpha = \frac{m_{\widetilde{B}}}{M_G}\,,  
\hspace*{1.0truecm} 
 z = \frac{m_{\rm mix}}{m_{\widetilde{B}}}\,,  
\hspace*{1.0truecm} 
 x = \frac{m_{\widetilde{X}}}{m_{\widetilde{B}}}\,.
\label{symbols}
\end{equation}
At the same level of precision in the small parameter 
$\alpha$, the reduced $3\times 3$ diagonalization matrix, 
defined by
\begin{equation}
  \left( \begin{array}{ccc}
    \tilde{B} & \tilde{X} & \tilde{\Psi} 
  \end{array} \right)^T 
 = D_{(3)} 
  \left( \begin{array}{ccc}
    \tilde{\chi}_0^0 & \tilde{\chi}_+^0 & \tilde{\chi}_-^0 
  \end{array} \right)^T \,,
\label{eigenvectors}
\end{equation}
is 
\begin{equation}
 D_{(3)} = 
  \left( \begin{array}{ccc}
  1- \alpha^2 \displaystyle{\frac{z^2}{2}}                           & 
  \displaystyle{
   \frac{z}{\sqrt{2}} \left[ \alpha +\alpha^2\! \left(1\!-\!\frac{x}{4}
                                              \right)\right] }       &
  \displaystyle{
   \frac{z}{\sqrt{2}} \left[ \alpha -\alpha^2\! \left(1\!-\!\frac{x}{4}
                                              \right)\right] }        
\\[2.5ex]
   - \alpha^2 z                                                      & 
  \displaystyle{
  \frac{1}{\sqrt{2}} \left[  1 +\alpha   \frac{x}{4}
                               -\alpha^2 \frac{x^2}{32} \right] }    &   
  \displaystyle{
  \frac{1}{\sqrt{2}} \left[ -1 +\alpha   \frac{x}{4}
                               +\alpha^2 \frac{x^2}{32} \right] }     
\\[2.5ex]   
   - \alpha z                                                        &  
  \displaystyle{
  \frac{1}{\sqrt{2}} \left[  1 -\alpha   \frac{x}{4}
                               -\alpha^2\! \left(
                   \frac{x^2}{32}\!+\!\frac{z^2}{2} \right) \right] }& 
  \displaystyle{
  \frac{1}{\sqrt{2}} \left[  1 +\alpha   \frac{x}{4}
                               -\alpha^2\! \left(
                   \frac{x^2}{32}\!+\!\frac{z^2}{2} \right) \right] }
\\[2.5ex]
 \end{array} \right) \,.
\label{redNmatrix}
\end{equation}

\section{Effective Dirac and Majorana Neutrino Masses}
\label{seesawformulae}
We are now in a position to evaluate Dirac and Majorana mass couplings
$m_D$ and $m_L$.  Since Majorana masses violate lepton number, they can
only be generated for nonvanishing $B M_R$.  In contrast, Dirac masses
can be obtained also for $M_R=0$.  In the following, we give expressions
for $m_D$ and $m_L$ classified according to the value of $M_R$. As in
the previous section, also here the flavor changing structure of the
soft parameters in the sneutrino potential is ignored. All the following
expressions for $m_D$ and $m_L$ hold for each generation. A
generalization to the case in which $A$, $B$, $m_{\,\tilde{l}}$, and
$m_{\tilde{\nu}}$, have a nontrivial matrix structure is technically
straightforward but leads to more involved formulas.

\subsection{Dirac neutrino mass $m_D$, $M_R \simeq 0$}
\label{diracmass-sterile}
\begin{figure}[ht]
\begin{center} 
\vspace*{0.3truecm}
\begin{picture}(100,80)(150,130)
  \ArrowLine(100,150)(150,150) \Text(125,160)[b]{$\nu_L$}
  \ArrowLine(200,150)(150,150) 
  \ArrowLine(200,150)(250,150) 
  \Line(195,145)(205,155) \Line(195,155)(205,145) 
  \Text(200,140)[t]{$\tilde{\chi}_j^0$} 
  \DashArrowArcn(200,150)(50,90,0){3}    \Text(155,195)[]{$\tilde{\nu}_L$}
  \DashArrowArcn(200,150)(50,180,90){3} \Text(245,195)[]{$\tilde{\nu}_R$}
  \DashArrowLine(200,230)(200,200){3}   \Text(215,215)[b]{$Av$}
  \ArrowLine(250,150)(300,150) \Text(275,160)[b]{$\nu_R$}
  \Vertex(150,150){1} 
  \Vertex(250,150){1} 
\end{picture}
\vspace*{0.3truecm}
\caption{Diagram contributing to the Dirac neutrino mass $m_D$.}
\label{Fig_diag}
\end{center}
\end{figure}
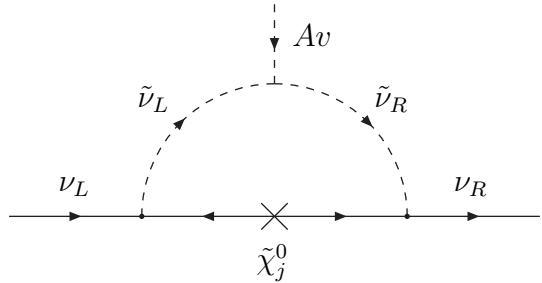

The radiatively generated Dirac mass $m_D$ is evaluated in the mass
eigenbasis for neutralinos and sneutrinos. The corresponding diagram is
shown in Fig.~\ref{Fig_diag}, where the sneutrino soft parameter $A$,
which breaks chiral symmetries, and the neutralino mass 
$m_{\tilde{\chi}_j^0}$, which connects left- and right-chiral fermions, 
are shown as mass insertions.  

The Majorana mass $M_R$, which couples the singlets $\bar{N}$, is
assumed to exist at the tree level, but to be small. Therefore, the
approximation in Eq.~(\ref{scalarssmatrix}) is sufficient for the
determination of the sneutrino mass eigenstates. They are, in this case,
only two, $\tilde{\nu}_1$ and $\tilde{\nu}_2$, and are both at the
electroweak scale. The current eigenstate vertex 
$\bar{\nu}_R \tilde{X} \tilde{\nu}_R$ has a coupling 
$-i \sqrt{2} g_X X_{\bar{N}}$ where $X_{\bar{N}}$ is the charge of the
singlet superfield ${\bar{N}}$ under $U(1)_{B-L}$. The gaugino
$\tilde{X}$ couples also to the left-handed neutrinos and sneutrinos
with a similar coupling $-i \sqrt{2} g_X X_L$.  A straightforward
calculation of the Dirac neutrino mass, in the one-generation case,
then, yields
\begin{equation}
m_D = 
  \frac{\left( A v \right)}{8 \pi^2} \sum_{j=1}^6 
  \left(g_X X_{\bar{N}} D_{2j} \right) 
    m_{\tilde{\chi}_j^0} 
  \left\{g_Y Y_L D_{1j} + \! g_X X_L D_{2j} 
                        + \! g_2 T_{3L} D_{6j} \right\}
  \,I(m^2_{\tilde{\nu}_1}, m^2_{\tilde{\nu}_2}, m^2_{\tilde{\chi}^0_j})\,,
\label{gener_dirmass}
\end{equation}
where the function $I(m^2_1, m^2_2, m^2_3)$ is defined in~\cite{BFPT},
$Y_L$ is the hypercharge of the SU(2) leptonic doublet, $T_{3L}$ is the
isospin of the neutral component of this doublet, and $D$ is the
neutralino diagonalization matrix defined in
Eq.~(\ref{neutreigenstates}).

This result becomes particularly transparent when the limit $v \to 0$ is
taken in the neutralino mass matrix. In this limit, the mixed
contribution of $\tilde{W}_3$ and heavy neutralinos vanishes, and the
third term in the curly bracket of Eq.~(\ref{gener_dirmass}) drops
out. Thus, in the approximation of vanishing phases in the neutralino
mass matrix, the Dirac neutrino mass becomes
\begin{eqnarray}
 m_D & \simeq & 
 \displaystyle{\frac{g_Y g_X }{8 \pi^2}}  \, Y_L X_{\bar{N}} 
 \left( A v\right)  m_{\rm mix}  
 \left\{ - \left(\frac{m_{\widetilde{B}}}{M_G}\right)^2 
        I(m^2_{\tilde{\nu}_1}, m^2_{\tilde{\nu}_2}, m^2_{\widetilde{B}})
 +      I(m^2_{\tilde{\nu}_1}, m^2_{\tilde{\nu}_2}, M^2_G)
 \right\}
\nonumber \\ 
  & & 
 +\displaystyle{\frac{g_X^2}{8 \pi^2}}  \, X_L X_{\bar{N}} 
 \left( \frac{A v}{m_{\widetilde{B}}}\right)  
\left(\frac{m_{\widetilde{X}}}{m_{\widetilde{B}}}\right) 
\left(\frac{m_{\widetilde{B}}}{M_G}\right)^2 \,,
\label{lim_mass}
\end{eqnarray}
where higher orders in the expansion parameter
$(m_{\widetilde{B}}/M_G)$ have been neglected.  Notice that in the
absence of a mass term mixing the $U(1)_Y$ and $U(1)_{B-L}$ gauginos, a
contribution to the Dirac mass $m_D$ comes from pure
$\tilde{X}$-$\tilde{X}$ exchange.  Finally, in the approximation
$ m^2_{\tilde{\nu}_1} \simeq  m^2_{\tilde{\nu}_2} \simeq 
  m^2_{\widetilde{B}} \ll M^2_G $, the above result can be cast in the 
simple form
\begin{equation}
 m_D  \sim  
 \displaystyle{\frac{g_X} {8\pi^2}}\,  X_{\bar{N}}  
 \left(\frac{A v }{m_{\widetilde{B}}}\right) 
 \left(\frac{m_{\widetilde{B}}}{M_G}\right)^2 
\left\{
 g_Y Y_L
 \left(\frac{m_{\rm mix}}{m_{\widetilde{B}}}\right)
 \left[ 
  2 \log\left(\frac{ M_G }{m_{\widetilde{B}}} \right)
  \!\! -\! \frac{3}{2} \right]
 +  g_X X_L 
 \left(\frac{m_{\widetilde{X}}}{m_{\widetilde{B}}}\right)
\right\}\,.
\end{equation}
For soft supersymmetry-breaking terms at the electroweak scale, the
Dirac neutrino mass $m_D$, besides being suppressed by a loop factor, is
moved away from $m_{\rm weak}$ by the ratio $(m_{\rm weak}/M_G)^2$. The
behavior $m_D \simeq (m_{\rm weak}/8\pi^2)(m_{\rm weak}/M_G)^2$ is quite
general and it is not limited to the choice of gauge group $U(1)_{B-L}$
made here. It has to be observed, however, that the suppression factor
$(m_{\rm weak}/M_G)^2$ is actually $(\tilde{m}/M_G)^2$, where
$\tilde{m}$ is a generic soft supersymmetry-breaking
parameter. Therefore, because of its dependence on a fixed scale $M_G$,
$m_D$ tends to grow in the superpartner decoupling limit. The
effectiveness of the suppression factor $(\tilde{m}/M_G)^2$ is recovered
only when also $M_G$ is increased and/or the trilinear parameters $A$
are decreased.

\subsection{Dirac neutrino mass $m_D$, $M_R \gg m_{\rm weak}$}
\label{diracmass-convent}
If the Majorana mass $M_R$ is non-negligible, then $CP$-even and
$CP$-odd components of the sneutrino states are split by lepton
flavor-violating mass terms. Therefore, when calculating $m_D$, the
general form in Eq.~(\ref{sneumassmatrix}) for the sneutrino mass
matrix has to be used, or at least that in Eq.~(\ref{snmass_mat_real})
obtained in the basis in which the $A$ and $B$ parameters are real. 
In this basis, $m_D$ gets the form
\begin{eqnarray}
m_D 
& = &  
  \frac{\left( A v \right)}{16 \pi^2} \sum_{j=1}^6 
  \left(g_X X_{\bar{N}} D_{2j} \right) 
    m_{\tilde{\chi}_j^0} 
  \left\{g_Y Y_L D_{1j} + g_X X_L D_{2j} 
                        + g_2 T_{3L} D_{6j} \right\}
\nonumber \\
&  &  
\hspace*{2truecm} \times 
         \left[ 
  I(m^2_{\tilde{\nu}_{+,\,1}}, m^2_{\tilde{\nu}_{+,\,2}}, 
                               m^2_{\tilde{\chi}^0_j}) + 
  I(m^2_{\tilde{\nu}_{-,\,1}}, m^2_{\tilde{\nu}_{-,\,2}}, 
                               m^2_{\tilde{\chi}^0_j})  
         \right]
 \,.
\label{gener_dirmass_2}
\end{eqnarray}
When taking
the limit $v\to 0$ in the neutralino mass matrix, also in this case, 
the mixed contribution of $\tilde{W}_3$ and heavy neutralinos vanishes.
Moreover, if $M_R \simeq M_G$, much above the electroweak scale, 
while 
$ m_{\tilde{\nu}_{-,\,1}} \simeq  m_{\tilde{\nu}_{+,\,1}} \simeq 
  m_{\,\tilde{l}}         \simeq  m_{\widetilde{B}}$, 
the above expression reduces to
\begin{eqnarray}
 m_D  & \sim  &
 \displaystyle{\frac{g_X} {8\pi^2}}\,  X_{\bar{N}}  
 \left(\frac{A v }{m_{\widetilde{B}}}\right) 
 \left(\frac{m_{\widetilde{B}}}{M_R}\right)^2 
\nonumber \\
&  &
\hspace*{1truecm} \times
\left\{
 g_Y Y_L
 \left(\frac{m_{\rm mix}}{m_{\widetilde{B}}}\right)
 \left[ 2 \log\left(\frac{M_R}{M_G}\right) \right]
 + g_X X_L 
 \left(\frac{m_{\widetilde{X}}}{m_{\widetilde{B}}}\right)
 \left[ 2 \log\left(\frac{M_R}{M_G}\right) -1 \right]
\right\} \,,
\end{eqnarray}
in the approximation of vanishing phases in the neutralino mass matrix.
Although different in some details, this expression for $m_D$ is
qualitatively very similar to that obtained when $M_R \simeq 0$. The
double suppression $(m_{\rm weak}/M_G)^2$ seems, therefore, a typical
feature of radiatively induced Dirac masses. It holds, indeed, for any
value of $M_R$.

\subsection{Majorana neutrino mass $m_L$, $M_R \gg m_{\rm weak}$}
\label{majoranamass}
\begin{figure}[ht]
\begin{center} 
\vspace*{0.3truecm}
\begin{picture}(100,80)(150,130)
 \ArrowLine(100,150)(150,150) \Text(125,160)[b]{$\nu_L$}
 \ArrowLine(200,150)(150,150) 
 \ArrowLine(200,150)(250,150) 
 \Line(195,145)(205,155) \Line(195,155)(205,145) 
 \Text(200,140)[t]{$\tilde{\chi}_j^0$} 
 \DashArrowArc(200,150)(50,0,45){3}    
               \Text(165,168)[]{$\tilde{\nu}_L$}
 \DashArrowArc(200,150)(50,45,90){3}   
               \Text(185,188)[]{$\tilde{\nu}_R$}
 \DashArrowArcn(200,150)(50,180,135){3}
               \Text(215,188)[]{$\tilde{\nu}_R$}
 \DashArrowArcn(200,150)(50,135,90){3} 
               \Text(235,168)[]{$\tilde{\nu}_L$}
 \DashArrowLine(200,230)(200,200){3} \Text(220,215)[b]{$B^\ast M_R$}
 \DashArrowLine(144.64,205.36)(164.64,185.36){3} 
               \Text(146.64,185.36)[b]{$A v$}
 \DashArrowLine(255.36,205.36)(235.36,185.36){3} 
               \Text(253.36,185.36)[b]{$A v$}
 \ArrowLine(300,150)(250,150) \Text(275,160)[b]{$\nu_L$}
 \Vertex(150,150){1} 
 \Vertex(250,150){1} 
\end{picture}
\vspace*{0.3truecm}
\caption{Diagram contributing to the Majorana neutrino mass $m_L$.}
\label{diag_Major}
\end{center}
\end{figure}
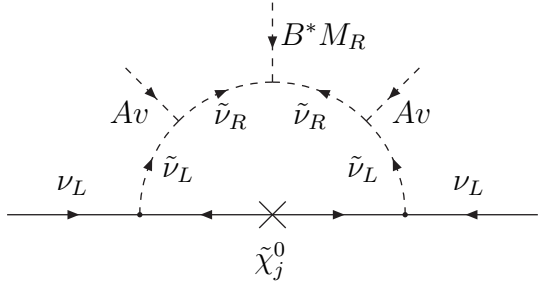

The Majorana neutrino mass $m_L$ violates lepton number and must have a
direct dependence on the soft parameter $B$, which is the only
lepton-number violating parameter in the sneutrino
potential.~\footnote{A different possibility was considered in the last
 reference in~\cite{BFPT}. In the absence of additional singlets
 $\bar{N}$, a lepton-number violating soft supersymmetry-breaking
 trilinear term was allowed. A mass splitting among left-handed
 sneutrinos was induced at the one-loop level and a Majorana neutrino
 mass $m_L$ at the two-loop level.}  
The diagram needed to obtain $m_L$ is shown in Fig.~\ref{diag_Major}. As
in the diagram of Fig.~\ref{Fig_diag}, the parameters $A$ and $B$ as
well as the relevant gaugino mass are shown as mass insertions, although
the calculation is done using the mass eigenstate formalism.
The neutrino mass is directly proportional to the splitting of each of
the two pairs of eigenvalues of the sneutrino mass
matrix~(\ref{snmass_mat_real}), induced by $B$. The largest contribution
to the Majorana mass comes from the splitting in the lightest
eigenvalues. The splitting in the two heavy eigenvalues, which appears
at a lower order in the expansion parameter $m_{\tilde{\,l}}/M_R$ than
the splitting in the two light eigenvalues, gives a much more suppressed
contribution. The resulting expression is
\begin{eqnarray}
 m_L  & = &   \frac{1}{4 \pi^2}  \sum_{j=1}^6
 \left(g_Y Y_L D_{1j} + \! g_X X_L D_{2j} 
                      + \! g_2 T_{3L} D_{6j} 
 \right) 
       m_{\tilde{\chi}_j^0}
 \left(g_Y Y_L D_{1j} + \! g_X X_L D_{2j}
                      + \! g_2 T_{3L} D_{6j} 
 \right) 
\nonumber  \\
      &   &    \hspace*{4truecm} 
\times  
  \left(\frac{A v}{m_{\tilde{\,l}}}\right)^2 \!
  \left(\frac{B}  {m_{\tilde{\,l}}}\right)   
  \left(\frac{m_{\tilde{\,l}}}{M_R}\right)^3   
 I(m_{\tilde{\nu}_{+,\,1}}^2, m_{\tilde{\nu}_{-,\,1}}^2, 
                              m^2_{\tilde{\chi}_j^0}) \,.
\label{maj_mass_gen}
\end{eqnarray}
When taking the limit $v \to 0$ in the neutralino mass matrix, the
contribution from mixed propagators $\tilde{B}$-$\tilde{W}_3$ drops out
and only the pure $\tilde{B}$-$\tilde{B}$, $\tilde{X}$-$\tilde{X}$, and
$\tilde{W}_3$-$\tilde{W}_3$ propagators survive. Furthermore, assuming a
real neutralino mass matrix, in the limit
$ m^2_{\tilde{\nu}_{+,\,1}} \simeq    m^2_{\tilde{\nu}_{-,\,1}} \simeq 
  m_{\tilde{\,l}}^2         \simeq  
  m^2_{\widetilde{B}}       \simeq    m^2_{\widetilde{W}}$, 
as well as $M_G \simeq M_R$ for the functions $I(.,.,.)$, the previous
expression becomes
\begin{equation}
 m_L   \simeq    
  \frac{1}{8 \pi^2}
  \left(\frac{A v}            {m_{\tilde{\,l}}^2}\right)^2 
  \left(\frac{B}              {m_{\tilde{\,l}}}  \right)
  \left(\frac{m_{\tilde{\,l}}}{M_R}              \right)^3  
\left\{
  \left(g_Y Y_L   \right)^2  m_{\widetilde{B}}
+ \left(g_2 T_{3L}\right)^2  m_{\widetilde{W}}
\right\} \,.
\label{maj_mass_lim}
\end{equation}
It is interesting to see that the suppression $(m_{\rm weak}/M_R)$ comes
with a higher power than in the expression for the Dirac mass $m_D$.  It
is, indeed, 
$m_L \sim (m_{\rm weak}/8\pi^2) (m_{\rm weak}\,\tilde{m}^2/M_R^3)$. Also
in this case, $m_L$ tends to increase in the superpartner decoupling
limit and its smallness can be retrieved by tuning the parameters $A$
and $B$ to be very small or raising the value of $M_R$.

\subsection{Majorana neutrino mass $m_L$, $M_R \ll m_{\rm weak}$}
\label{majmass_smallMR}
If $M_R$ is much smaller than the electroweak scale, the Majorana
neutrino mass $m_L$ is not suppressed by the large scale $\sim M_G$.  In
this case, all particles exchanged in the loop in Fig.~\ref{diag_Major}
have masses of the order of the electroweak scale, and they all
contribute to $m_L$.  Thus, the suppression factor only comes from
$(M_R/m_{\rm weak})$ when all the supersymmetry-breaking parameters are
around the weak scale.  The Majorana neutrino mass is given by
\begin{eqnarray}
 m_L  & = &   \frac{1}{4 \pi^2}  \sum_{j=1}^6
 \left(g_Y Y_L D_{1j} + \! g_X X_L D_{2j}
                      + \! g_2 T_{3L} D_{6j}
 \right)
       m_{\tilde{\chi}_j^0}
 \left(g_Y Y_L D_{1j} + \! g_X X_L D_{2j}
                      + \! g_2 T_{3L} D_{6j}
 \right)
\nonumber  \\
\nonumber \\
&  &  
\times (B M_R)
\frac{(A v)^2}{(m^2_{\tilde{\nu}_2}-m^2_{\tilde{\nu}_1})^2}
         \left[ 
  I(m^2_{\tilde{\nu}_1}, m^2_{\tilde{\nu}_1}, 
                         m^2_{\tilde{\chi}^0_j}) + 
  I(m^2_{\tilde{\nu}_2}, m^2_{\tilde{\nu}_2}, 
                         m^2_{\tilde{\chi}^0_j}) - 
 2I(m^2_{\tilde{\nu}_1}, m^2_{\tilde{\nu}_2}, 
                         m^2_{\tilde{\chi}^0_j})  
         \right]
\,,
\label{maj_mass_ext}
\end{eqnarray}
where $m_{\tilde{\nu}_{1,2}}^2$ are defined in Eq.~(\ref{eigen_nu}).
When taking the limit $v \to 0$ in the neutralino mass matrix, the
contribution from mixed propagators $\tilde{B}$-$\tilde{W}_3$ drops out
and the dominant contributions come from the $\tilde{B}$-$\tilde{B}$ and
$\tilde{W}_3$-$\tilde{W}_3$ propagators.  It is easy to see from
Eq.~(\ref{maj_mass_ext}) that the Majorana neutrino mass $m_L$ generated
when $M_R \ll m_{\rm weak}$ is of order $(1/8\pi^2) (B M_R/m_{\rm
weak})$.

The expression~(\ref{maj_mass_ext}) is valid also for the radiative
contribution to $m_L$ obtained in the second (tree-level) scenario of
Sec.~\ref{mechanism-1}, in which no $U(1)_{B-L}$ symmetry is imposed,
and generically for the contributions obtained when $M_R \ll m_{\rm
weak}$ and $B M_R \lesssim m_{\rm weak}^2 $. As was already argued in
Sec.~\ref{sneutrinomm}, acceptable values of $m_L$ arise, in general,
when $B M_R$ is not too close to $m_{\rm weak}^2$. Although the
supersymmetry-breaking parameter $B$ is very large, the approximations
made in Sec.~\ref{sneutrinomm}, as well as the resulting
expression~(\ref{maj_mass_ext}), are still applicable.

\section{Effective Yukawa Neutrino and Sneutrino Interactions}
\label{effectiveinteractions}
Yukawa interaction terms Higgs boson--lepton doublet--right-handed
neutrino are also induced radiatively. The diagrams for the neutral
Higgs boson--left-handed-neutrino--right-handed-neutrino interaction
term are similar to that in Fig.~\ref{Fig_diag}, with the external
scalar line being that of a physical neutral Higgs boson.  Analogous
diagrams generate also the vertex of charged Higgs bosons--charged
leptons--right handed neutrinos. The interaction terms ${\nu_L} \nu_L
H^0 H^0$ as well as $e e H^+ H^+$ are obtained from variations of the
diagram in Fig.~\ref{diag_Major}. The main characteristic of the
couplings associated with these terms is that they are momentum
dependent with, in general, complicated form factors. Although different
from $m_D/v$ and $m_L/v^2$, they have, however, the dimensionality and
order of magnitude of these ratios.

For a Dirac mass in the eV range, for example, the effective Yukawa
coupling is $\sim 10^{-11}$. The smallness of this coupling is now {\it
natural} and, as the smallness of $m_D$, owes its origin to the seesaw
pattern of widely different scales for the neutralinos entering in the
loop calculation. In the SM, when the Higgs mechanism is the only way to
obtain small neutrino masses, such a small value can only be imposed by
hand. In the traditional seesaw mechanism, the couplings for Yukawa
interaction terms are of ${\cal O}(1)$.  While the phenomenological
relevance of these terms for possible collider signals is marred by the
presence of right-handed fields at scales $\sim 10^{12}\,$GeV, these
large couplings have subtle consequences in other sectors. For example,
they alter the pattern of Yukawa couplings
unification~\cite{YUKUNIF}. Moreover, they lift the predictions for
lepton-flavor violating processes involving charged leptons, such as the
decay $\mu \to e \gamma$, by inducing sizable splittings among charged
sleptons of different generations. Indeed, these processes, which are
realized at the quantum level, may be detectable if the mass of the
charged sleptons exchanged in the loops is not too heavy~\cite{LFV}.

In the scenarios presented here, with intrinsically small effective
Yukawa couplings, large rates for lepton-flavor-violating processes are
obtained if a large splitting among the charged-slepton masses of
different generations exists already at the Planck scale or at the scale
of $U(1)_{B-L}$ breaking.  Notice that, although small, these effective
Yukawa couplings, obtained radiatively, are, in general, much larger
than those obtained at the tree level from the operators in
Eq.~(\ref{softA}). Indeed, these tree-level Yukawa couplings are
naturally $\sim 10^{-13}$; see the discussion in Secs.~\ref{mechanism-1}
and~\ref{models}.  In some specific cases, however, they may be larger
and even comparable to those obtained radiatively.  In this case, the
actual couplings of neutrino Yukawa interactions are of mixed origin.
They remain, however, very small.

The dimensionful coupling $m_L/v^2$ for the interaction terms 
${\nu_L} \nu_L H^0 H^0$ and $e e H^+ H^+$ is 
$\sim (10^{13}\,$GeV$)^{-1}$ for $m_L \sim 1\,$eV. This is the same
dimensionful suppression factor that can be obtained by integrating out
right-handed neutrinos of mass $M_R \sim 10^{13}\,$GeV if these couple
at the tree level to Higgs and lepton doublets with couplings of 
${\cal O}(1)$. Couplings that may
arise from operators such as those in Eq.~(\ref{majmassleft}), in the
case of a gauged $U(1)_{B-L}$ symmetry or from operators such as the
first in Eq.~(\ref{m_nu_spotential}) when $U(1)_{B-L}$ is not
introduced, are much smaller than the effective couplings obtained 
radiatively.

Similar considerations hold also for Higgsino-sneutrino-neutrino
operators, which also arise at the quantum level~\cite{BFPT}. Their
couplings are, in general, so small as to be irrelevant for
phenomenology.

Large are, on the contrary, the trilinear sneutrino soft terms $A$,
which can be as large as the electroweak scale.  In this case,
relatively large interaction terms 
$ A H^0 \, \tilde{\nu}_R^\ast \,\tilde{\nu}_L $ and 
$ A H^+ \, \tilde{\nu}_R^\ast \, \tilde{e}_L $ are present. They may
possibly affect searches for neutral and charged Higgs bosons.
Particularly interesting is, in this respect, the situation in which the
right-handed sneutrino $\tilde{\nu}_R$ is very light. (For the effect of
light sneutrinos in searches of the charged Higgs boson, within a
generic supersymmetric model, see~\cite{BD}.)  A dedicated study of this
possibility is certainly important and is left for future work.

\section{Physical Neutrino Masses}
\label{neumasses}
It was shown in the previous sections that small neutrino masses can be
naturally obtained in supersymmetric models with right-handed neutrinos,
in which (i) the lowest order neutrino Yukawa operators are forbidden;
(ii) effective Yukawa couplings are induced through higher-dimensional
operators by a spurion field $Z$ with a supersymmetry-conserving VEV
much smaller than the supersymmetry-violating one, and (iii) a
$U(1)_{B-L}$ gauge symmetry may be introduced.  In this class of models,
contributions to Dirac and Majorana neutrino masses $m_D$ and $m_L$
entering the neutrino mass matrix
\begin{equation}
\left( 
 \begin{array}{cc} 
     m_L & m_D  \\ m_D & M_R 
 \end{array} 
\right)  
\end{equation} 
are generated at the tree level and/or at the quantum level. The overall
results obtained for neutrino masses can be classified according to the
value of the right-handed neutrino mass $M_R$.

\noindent (1) 
If $M_R$ is large enough to be hierarchically split from $m_{\rm weak}$,
$M_R \gg m_{\rm weak}$, the complete or largest contributions to $m_L$
and $m_D$ are, naturally, of radiative origin if the light neutrino
states are assumed to be $\sim 1\,$eV. This case is naturally realized
when a $U(1)_{B-L}$ gauge symmetry broken at a large scale $M_G$ is
introduced.  If $M_R \simeq M_G$, the Majorana mass $m_L$, with its
dependence $m_L \sim (m_{\rm weak}/8\pi^2)(m_{\rm weak}/M_R)^3$, is much
smaller than the Dirac mass $m_D$, which is 
$\sim (m_{\rm weak}/8\pi^2)(m_{\rm weak}/M_G)^2$.  Nevertheless, the
eigenstates $\nu_i$ have mass dominated by $m_L$. The eigenstates $n_i$
are, as in the usual seesaw mechanism, at the large scale $M_R$. The six
eigenstates are Majorana fermions. The mixing angle $\theta_\nu$ between
light and heavy states, is 
$\vert \sin 2\theta_\nu \vert \sim 2 \vert m_D \vert /M_R$. Therefore,
the states $\nu_i$ and $n_i$ are, respectively almost pure active and
sterile neutrinos. Light states $\nu_i$ at the eV scale require 
$M_R (\sim M_G) \sim 100\,$TeV. This scale, however, can be decreased if
the soft parameters $A$ and $B$ are suppressed with respect to 
$m_{\rm weak}$. Notice that for light neutrinos in the sub-eV range, the
tree-level contribution to $m_D$ may be non-negligible with respect to
the radiative one and affect the mixing angle.

\noindent (2) 
If $M_R$ is around the electroweak scale, $m_L$ is, in general,
suppressed with respect to $m_{\rm weak}$ only by numerical loop
factors. A drastic tuning of $A$ and/or $B$ is then required to reduce
$m_L$ to the eV scale.  [Small values of these parameters may naturally
occur in a gauge-mediated supersymmetry-breaking scenario (see
Sec.~\ref{models}).]  Also this case, as the previous one of large
$M_R$, can be realized when a $U(1)_{B-L}$ gauge symmetry is introduced.

\noindent (3)
If $M_R$ is much smaller than the electroweak scale, the radiative
contribution to $m_L$ is given by $\sim (1/8\pi^2) (BM_R/m_{\rm weak})$.
The Dirac mass $m_D$ has a tree-level contribution 
$\sim {\cal A}_Z v/M_P$ with $v$ an electroweak VEV.  In addition, if no
$U(1)_{B-L}$ symmetry is introduced, the left- and right-handed Majorana
masses $m_L \sim v^2/M_P$ and $M_R \sim {\cal A}_Z^2/M_P$ are also
generated at the tree level as discussed in the second scenario of
Sec.~\ref{mechanism-1}.  If a $U(1)_{B-L}$ gauge symmetry is introduced,
on the other hand, we have a radiative contribution to the Dirac mass of
$\sim (m_{\rm weak}/8\pi^2) (m_{\rm weak}/M_G)^2$.  The relative size of
$M_R$ with respect to that of $m_D$ has actually rather important
physical consequences. Therefore, we further distinguish the following
two cases.

\noindent (3a) 
For $M_R \gtrsim m_D$, the six Majorana eigenstates $(\nu_i,n_i)$ have
masses $\sim m_L+m_D^2/M_R$, $M_R$, if there is a hierarchy between
$M_R$ and $m_D$, or $(1/2) \{ M_R \pm \sqrt{M_R^2-4 m_D^2}\}$, if $M_R$
is comparable to $m_D$. The first case can be realized through the
tree-level or the radiative mechanism, or both; the second is typical of
the radiative generation of neutrino masses with a $U(1)_{B-L}$ gauge
symmetry.  The mixing angle between $\nu_i$ and $n_i$ ranges from small,
for $M_R > m_D$, to maximal, for $M_R \sim m_D$. Notice that $M_R \sim
10^{-3}\,$eV and $m_D \sim 10^{-5}\,$eV could explain the solar neutrino
problem with an oscillation $\nu_e \to \nu_s $ and a small mixing angle. 
In order to avoid the same pattern of oscillation also in the other two
generations, a particular flavor structure of the trilinear $A$ terms
and/or of the different $M_R$ has to be implemented.

\noindent (3b) 
The case $M_R \ll m_D$ is typically realized in the radiative mechanism
of neutrino Dirac mass generation. In this case, almost degenerate
Majorana states $(\nu_i, n_i)$ are obtained with a nearly maximal
mixing. Their masses are of order 
$m_D \sim (m_{\rm weak}/8\pi^2) (m_{\rm weak}/M_G)^2$, so that a scale
of $B\!-\!L$ violation of $\sim 10^{7}\,$GeV is required to obtain
neutrino masses in the eV range.

\noindent (4)
Finally, there is the possibility $M_R = 0$, for which the Majorana
states $(\nu_i, n_i)$ have masses $\pm m_D$ and combine into a Dirac
neutrino, for each family.  As discussed in the previous sections, in
both mechanisms of neutrino-mass generation, this possibility can be
realized by imposing an additional symmetry, such as a lepton-number
symmetry or its discrete subgroup given in Eq.~(\ref{lepton-num}).
Notice that purely Dirac neutrinos emerge in this case, even when a
$U(1)_{B-L}$ gauge symmetry is imposed.  From the phenomenological point
of view this case is very different from all the others discussed above. 
The observed oscillation structure of atmospheric and solar oscillations
can then be explained only as a flavor oscillation. It is obviously very
difficult to accommodate in such a scenario the results from the LSND
experiment.  Naturally, no signal is expected from the $\beta \beta$
decay. Moreover, relatively large neutrino dipole moments are predicted,
i.e., dipole moments which are not loop suppressed with respect to
neutrino masses (see the discussion on quark and lepton dipole moments
in a scenario of radiative generation of fermion masses~\cite{BFPT}).

Both mechanisms for generating small neutrino masses proposed in this
paper offer a very fertile ground in which widely different scenarios of
neutrino masses and mixing can be implemented. If the lowest order
tree-level Yukawa couplings exist, then the loops calculated above
provide corrections to the tree-level neutrino masses.  For unsuppressed
Yukawa couplings, i.e., for couplings of ${\cal O}(1)$, these
corrections are negligibly small once the scale $M_R$ is raised, as in
the usual seesaw case, to get tree-level masses of the correct order of
magnitude.  If tree-level Yukawa couplings are, however, suppressed by
some other mechanism, as explained in Secs.~\ref{mechanism-1}
and~\ref{models}, then a large variety of neutrino spectra is
possible. The fact that the lowest-order neutrino Yukawa operators are
vanishing is therefore a key ingredient to unravel a much wider range of
possibilities than that offered by the traditional seesaw mechanism.

\section{Model Embedding}
\label{models}
In this section, an explicit model that generates the desired hierarchy
between the supersymmetry-conserving and the supersymmetry-violating VEV
of the $Z$ field, i.e., ${\cal A}_Z \ll {\cal F}_Z$, is presented.  We
recall that ${\cal A}_Z $ is responsible for the tree-level generation
of neutrino masses, whereas ${\cal F}_Z$ gives rise to the sneutrino
soft trilinear $A$ parameter. This, together with a neutralino mass,
induces quantum contributions to the Dirac mass $m_D$. The same
parameter $A$, together with the soft bilinear $B$ parameter and a
neutralino mass, induces a radiative contribution to the Majorana mass
$m_L$. Comments on the size of the parameters $A$ and $B$ in various
supersymmetry-breaking scenarios are also made.

The model is based on a supersymmetric SU(2) gauge theory with four
doublet chiral superfields $Q_i$ and six singlet superfields 
$Z^{ij} = - Z^{ji}$ where $i,j$ are flavor indices
($i,j=1,\cdots,4$)~\cite{IYIT}.  All the $Q_i$ and $Z^{ij}$ fields are
singlets under the SM gauge groups.  Without a superpotential, this
theory has a flavor SU(4)$_F$ symmetry.  This is explicitly broken to a
global SP(4)$_F$ subgroup if the tree-level superpotential~\cite{INTY}
\begin{eqnarray}
  W_{\rm tree} = \lambda Z (QQ) + \lambda_Z Z^a (QQ)_a 
\label{tree-sup}
\end{eqnarray}
is introduced. Here, $(QQ)$ and $Z$ are singlets of $SP(4)_F$, while
$(QQ)_a$ and $Z^a$ are five-dimensional representations of $SP(4)_F$
$(a=1,\cdots,5)$.  [$Z$ and $Z^a$ are certain linear combinations of the
$Z^{ij}$ fields; $(QQ)$ and $(QQ)_a$ combinations of the gauge-invariant
operators $(Q_i Q_j)$.]  The model has a discrete symmetry $Z_n$ that
assigns the following charges to the $Q_i$ and $Z^{ij}$ fields:
\begin{eqnarray}
  Z_n \left(Q_i\right) = \frac{1}{2}, \quad
  Z_n \left(Z^{ij} \right) = -1 \,,
\end{eqnarray}
and that is identified with the symmetry used in Secs.~\ref{mechanism-1}
and~\ref{newseesaw} to forbid tree-level neutrino Yukawa terms 
$\bar{N} L H$.~\footnote{Here, it is assumed that quadratic (if $n=2$)
and cubic (if $n=3$) terms in $Z$ are absent.}  Then, the $Z$ field can
couple to the SM fields through the superpotential in Eq.~(\ref{softA}).

The dynamics of the SU(2) gauge theory causes the condensation of the
$Q_i$ fields through nonperturbative effects~\cite{IS}.  By integrating
out the SU(2) gauge fields together with $Q_i$ and $Z^a$, the low-energy
effective superpotential
\begin{eqnarray}
  W_{\rm eff} \simeq \lambda \Lambda^2 Z
\label{lin_sup}
\end{eqnarray}
is obtained for $\lambda_Z > \lambda$, where $\Lambda$ is a dynamical
scale of the SU(2) gauge interaction. [Notice that there is a mixed
$Z_n$--SU(2) anomaly and that $\Lambda$ has therefore a $Z_n$ charge.]
Thus, the $Z$ field acquires a nonvanishing $F$ term, 
${\cal F}_Z \simeq \lambda \Lambda^2 \neq 0$~\cite{IYIT}, which induces
a sneutrino $A$ parameter of order ${\cal F}_Z/M_P$.

The $Z$ direction is flat at the tree level, but it is lifted by loop
corrections.  The corrections to the K\"ahler potential from the strong
SU(2) gauge interaction are noncalculable and can only be
estimated. They give rise to a quadratic term in $Z$ in the scalar
potential. If positive, this term, together with a linear term due to
supergravity effects, induces a tiny VEV for the scalar component of
$Z$~\cite{KIITY}.  However, there are also contributions to the mass
squared for the $Z$ field near the origin coming from loops of light
particles. These are calculable and larger than the noncalculable ones
as long as $\lambda$ is in a perturbative regime~\cite{CLP}.  The sign
of the mass squared is shown to be positive~\cite{CLP}, so that $Z = 0$
is a local minimum of the potential in the limit of a global
supersymmetry.  Thus, the scalar potential for the $Z$ field has the
form
\begin{eqnarray}
  V \simeq \frac{|\lambda|^4}{16\pi^2} \Lambda^2 |Z|^2
    + \left( \lambda \Lambda^2 m_{3/2} Z + {\rm h.c.} \right)\,,
\end{eqnarray}
and the nonvanishing VEV for the scalar component of the $Z$ field is
\begin{equation}
{\cal A}_Z \simeq \frac{16\pi^2}{\lambda^3} m_{3/2}  \,.
\end{equation}
This VEV, in turn, induces effective tree-level neutrino Yukawa
couplings of order ${\cal A}_Z/M_P$ through the superpotential term in
Eq.~(\ref{softA}).  If $m_{3/2} \simeq 1\,$TeV and $\lambda = O(1)$, the
resulting Dirac neutrino masses, $m_D \simeq 10^{-2}\,$eV, are
sufficiently large to explain the solar neutrino deficit by the MSW
mechanism.  If $\lambda$ is somewhat smaller than 1, it is even possible
to generate neutrino masses in the range needed by LSND and atmospheric
neutrino experiments. Note that this tree-level generation of neutrino
masses generically occurs if a chiral superfield $Z$, which couples to
SM fields through the superpotential term in Eq.~(\ref{softA}), acquires
a nonvanishing auxiliary component at the tree level, but has a
vanishing scalar component in the global supersymmetry limit.

Thus, the above model can yield the desired size of the $A$ parameter
and of the effective tree-level Yukawa couplings through the
supersymmetry-violating and supersymmetry-conserving VEV's of the $Z$
field.  On the other hand, a $B$ parameter of the order of the gravitino
mass $m_{3/2}$ is always generated by supergravity effects.  In
addition, if the $Z_n$ symmetry needed to forbid the renormalizable
neutrino Yukawa operators is generic (i.e., $n \neq 2$), the $B$
parameter also has a contribution of order ${\cal F}_Z/{\cal A}_Z$ from
Eq.~(\ref{majmassuno}), which is generically much larger than $m_{3/2}$.
(In this case, the value of $M_R$ is, however, rather small.)  To
summarize, in the present model, we obtain the $A$ and $B$ parameters
\begin{equation}
  A \simeq \frac{\Lambda^2}{M_P}\,,   \hspace*{1truecm}
  B \simeq \left\{ \begin{array}{ll}
 m_{3/2} + \displaystyle{\frac{\Lambda^2}{16\pi^2 m_{3/2}}} 
             & \quad [{\rm for}\, Z_n \, (n \neq 2)]\,, \\[1.5ex] 
 m_{3/2}     & \quad [{\rm for}\, Z_2]\,, 
    \end{array} \right. 
\label{AandB}
\end{equation}
and the tree-level neutrino Yukawa couplings $y_{\nu}$:
\begin{equation}
  y_{\nu} \simeq  \frac{16\pi^2 m_{3/2}}{M_P}\,.
\label{general_Yukawa_Z}
\end{equation}
Here, we have set $\lambda$ and the coupling of the superpotential in
Eq.~(\ref{softA}) to be of order 1.  For simplicity, only the case of $B
\simeq m_{3/2}$ will be considered in the following
discussion.~\footnote{If the
 discrete symmetry in Sec.~\ref{newseesaw} is generic, i.e., $Z_n$ with
 $n \neq 2$, it is possible to have $M_R \simeq 0$ and
 $M_R B \mathop{}_{\textstyle \sim}^{\textstyle <} m_{\rm weak}^2$.
 Then, the neutrino mass matrix takes the form
 $\sim ((m_L, m_D)^T, (m_D, 0)^T)$, where $m_L$ is generated by the
 diagram shown in Fig.~\ref{diag_Major}, and it is suppressed with
 respect to $m_{\rm weak}$ only by a numerical loop factor if
 $A^2 \simeq (M_R B) \simeq m_{\rm weak}^2$.}

It is interesting to compare the size that the above parameters acquire
in supergravity-mediated~\cite{GRAVITY} and
gauge-mediated~\cite{GAUGE1,GAUGE2} supersymmetry-breaking scenarios. In
the first, the gravitino mass $m_{3/2}$, as well as the $B$ parameter,
is at the electroweak scale $\sim 1\,$TeV.  The effective neutrino
Yukawa couplings are of order $10^{-13}$ and induce Dirac neutrino
masses of order $m_D \simeq 10^{-2}\,$eV.  On the other hand, the $A$
parameter is less constrained. Since the $Z$ field is charged under the
$Z_n$ symmetry needed to forbid the lowest order Yukawa operators, its
VEV ${\cal F}_Z$ cannot give sizable gaugino masses.  Therefore, an
additional singlet $S$ with a supersymmetry-violating VEV 
${\cal F}_S \simeq m_{3/2} M_P$ is needed to give soft
supersymmetry-breaking mass to all superparticles.  Thus, the
phenomenological constraint on the supersymmetry-violating VEV of the
$Z$ field, ${\cal F}_Z$, is only 
${\cal F}_Z \mathop{}_{\textstyle \sim}^{\textstyle <} {\cal F}_S$, and
this translates into 
$A \mathop{}_{\textstyle \sim}^{\textstyle <} m_{3/2}$.  Although it is
logically possible that the $A$ parameter is much smaller than the
electroweak scale, we may naturally expect that ${\cal F}_Z$ and 
${\cal F}_S$ are somehow related to each other.  If this is the case,
the $A$ parameter is also of order of the gravitino mass. This situation
was assumed throughout the previous sections.

In a gauge-mediated scenario, the VEV ${\cal F}_Z$ of the $Z$ field can
be the primary source of supersymmetry breaking. The
supersymmetry-breaking effect in the $Z$ field can be transmitted to all
the superparticles through gauge interactions~\cite{GAUGE1,GAUGE2,NTY},
and no additional singlet is necessary.  In this case, both $A$ and $B$
parameters are of the order of the gravitino mass (i.e., $\Lambda^2
\simeq m_{3/2} M_P$).  Since in the gauge-mediated scenario $m_{3/2}$
is, in general, much smaller than that in the supergravity-mediated one,
the $A$ and $B$ parameters are certainly below the electroweak
scale. Thus, the neutrino Yukawa coupling in
Eq.~(\ref{general_Yukawa_Z}) cannot give neutrino masses in the range of
the various $\Delta m^2$ suggested by different experiments, except
perhaps for the vacuum-oscillation region of the solar neutrino
experiments.  A substantial contribution to neutrino masses may come
from the radiative mechanism.  Notice that all the expressions for $m_L$
and $m_D$ derived in Sec.~\ref{seesawformulae} still apply to the
gauge-mediated case.  However, since the values of $A$ and $B$ are now
smaller, the values of $M_R$ and of the ($B-L$)-breaking scale $M_G$
needed to obtain sizable neutrino masses are also lower than those
needed in the supergravity-mediated case.

It should be noticed here that, if the $Z_n$ charge $-1$ is assigned not
only to the $\bar{N}$ field but also to all SU(2)-singlet SM fields
$\bar{U},\bar{D}$, and $\bar{E}$, then all the SM Yukawa couplings are
forbidden and effective Yukawa operators are generated also for quark
and charged-lepton fields through higher-dimensional operators such as
those in Eq.~(\ref{softA}). The tree-level contributions that these
operators give to Yukawa couplings of quarks and charged leptons are,
however, completely negligible. Thus, their masses, as well as Yukawa
couplings, are generated radiatively, as in the framework of
Ref.~\cite{BFPT}, through trilinear supersymmetry-breaking parameters
induced by ${\cal F}_Z$, and gluino and/or neutralino masses. In this
case, a unified radiative generation picture for quark and lepton masses
emerges. The smallness of neutrino masses with respect to the masses of
all other fermions is then easily explained by the fact that neutrinos
feel the presence of neutralinos and sneutrinos at the large scales
$M_G$ and $M_R$ of $U(1)_{B\!-\!L}$ violation.  In contrast, the leading
contribution to quark and charged-lepton masses comes from the exchange
of sfermions and gluinos or sfermions and neutralinos at the electroweak
scale.

\section{CONCLUSIONS}
\label{conclusion}
In this paper, two different mechanisms to obtain light physical
neutrino states in supersymmetric models with three right-handed
neutrinos are proposed. It is observed that if the lowest order neutrino
Yukawa operators are forbidden by means of a horizontal discrete
symmetry, Yukawa couplings are still induced by higher-dimensional
operators suppressed by the Planck mass. In these operators, the three
fields that participate in a Yukawa interaction are coupled to a spurion
field $Z$, which acquire a supersymmetry-conserving and a
supersymmetry-violating VEV, respectively, ${\cal A}_Z$ and 
${\cal F}_Z$. If a large hierarchy between the two VEV's is possible,
then naturally small Yukawa couplings can be induced at the tree level.

Models where such a hierarchy can be implemented exist and one is
explicitly described in this paper. In this model, the $Z$ direction is
flat at the tree level. This flatness is lifted by loop corrections, and
a supersymmetry-conserving VEV ${\cal A}_Z \sim 16 \pi^2 m_{3/2} $ is
induced by supersymmetry breaking.  The strong suppression of neutrino
Yukawa couplings $y_{\nu}$ gets naturally linked to the hierarchy
between $m_{3/2}$ and $M_P$: 
$y_{\nu} \sim {\cal A}_Z/M_P \sim 16 \pi^2 (m_{3/2} /M_P)$. If Majorana
masses for active and sterile neutrinos are forbidden by an additional
symmetry such as, for example, lepton number, only Dirac neutrino masses
are generated. They are of order $\sim {\cal A}_Z v /M_P $, with $v$ an
electroweak VEV, i.e., they can reach $\sim 10^{-2}\,$eV if all
interaction couplings are assumed of ${\cal O}(1)$. Intergenerational
mass splittings and related oscillations rely on specific textures of
these couplings.

Nevertheless, it is possible to forbid only the lowest order mass term
for right-handed neutrinos, but leave allowed higher-dimensional mass
operators for left- as well as right-handed neutrinos (see the second
tree-level scenario discussed in Sec.~\ref{mechanism-1}). Besides the
Dirac mass $m_D$ given above, also Majorana masses for active and
sterile neutrino $m_L$ and $M_R$ are present, respectively of order
$\sim v^2 /M_P$ and $\sim {\cal A}_Z^2 /M_P$. Six physical states of
Majorana type are generated: $\nu_i$ and $n_i$ ($i=1,2,3$).  The states
$n_i$ correspond mainly to sterile neutrinos and are the heaviest
ones. For all interaction couplings of ${\cal O}(1)$, their mass is
$10^4$--$10^6$ times larger than the mass of the states $\nu_i$, which
is of order $\sim 10^{-5}\,$eV. Different patterns for different
generations may be induced by allowing a nontrivial structure of the
interaction couplings.

The second mechanism proposed is that of the radiative generation of
neutrino masses via sneutrino-neutralino loops.  It is illustrated in
the supersymmetrized version of a typical model that leads to the
traditional seesaw mechanism: i.e., a supersymmetric model with three
right-handed neutrinos and an additional gauge interaction broken at a
large scale $M_G$. Nevertheless, large radiative contributions to the
Majorana mass $m_L$ may arise also in models without any additional
gauge interaction, but with an explicit violation of lepton number.
Also for this mechanism, as in the tree-level one, the chief assumption
is the vanishing of the lowest-order neutrino Yukawa operators.

It is shown that (i) the Majorana neutrino mass $M_R$ for right-handed
neutrinos is not necessarily of order ${\cal O}(M_G)$ and can even be
exactly vanishing, that (ii) a seesaw pattern of light-heavy scales is
present in the neutralino and (possibly) the sneutrino mass matrix.
Dirac and Majorana masses $m_D$ and $m_L$ are induced by trilinear and
bilinear soft sneutrino terms as well as neutralino mass terms, which
provide, respectively, information on chirality breaking, on
lepton-number violation and on the correct number of $R$ charges. Apart
from a loop suppression factor, ratios of light over heavy scales,
present in the neutralino and sneutrino mass matrices, steer $m_D$ and
$m_L$ away from the electroweak scale. Since these ratios enter in the
expression for $m_D$ and $m_L$ with powers larger than 1, their
effectiveness in suppressing $m_D$ and $m_L$ is stronger than in the
usual seesaw mechanism, and the scale $M_G$ can be much lower.

Such a mechanism allows a large variety of possible neutrino spectra. In
the case of exact vanishing of $M_R$, three Dirac neutrino states are
possible, as in the first scenario of the tree-level mechanism or in a
traditional Higgs mechanism with highly tuned Yukawa couplings.  These
states can have now arbitrary mass depending on the value of $M_G$.  In
this case, a texture of neutrino masses consistent with experimental
observations finds its origin in the texture of the trilinear $A$ terms
for sneutrinos.

In general, for nonvanishing $M_R$, six Majorana states $\nu_i$ and
$n_i$, are generated.  For heavy $M_R$, the three states $n_i$ are heavy
and no space is left for light sterile neutrinos. For light $M_R$, the
situation is very rich.  The six states are all light. The three $n_i$
neutrinos, however, can still be much heavier than the $\nu_i$'s, in
which case the mixing angles between $n_i$'s and $\nu_i$'s are very
small.  There are finally the other interesting cases in which the
$\nu_i$ and $n_i$ states are roughly at the same scale or nearly
degenerate, with mixing angles ranging from small to maximal.  Flavor
oscillations can be accommodated in this class of scenarios, as before,
i.e., relying on specific textures of the trilinear and bilinear
neutrino soft parameters.  Oscillations among active and sterile
neutrinos are, however, also possible. Whether this is the answer to the
puzzle posed by the present results of solar, atmospheric, and reactor
neutrino experiments is a question which new experimental data will
answer, hopefully, soon.  Irrespective of this, it is nontrivial that
such oscillations, as well as those between active and much heavier
sterile neutrinos advocated in supernova physics, can be easily
accommodated in this class of scenarios.

Since neutrino masses are much smaller than all other masses, it is
plausible to assume that the lowest-order neutrino Yukawa operators
inducing neutrino masses are vanishing and that the origin of neutrino
masses is linked to operators of higher dimensionality or to radiative
generation. We have shown that both possibilities are easily implemented
in supersymmetric models.  The resulting mechanisms encompass the
typical Higgs and seesaw mechanisms, although with different
realizations, and give also rise to various interesting neutrino
spectra.

\acknowledgements 

The authors thank T.~Yanagida for discussions and interest in this
work. They also acknowledge discussions with S.~Davidson and
K.~Hagiwara.  F.B. is supported by a Japanese Monbusho fellowship and
thanks the KEK theory group, in particular Y.~Okada, for
hospitality.  The work of Y.N. was supported by the Japanese Society
for the Promotion of Science.

{\bf Note added}
While this work was being completed, we received Ref.~\cite{AHMSW}, 
which makes another proposal for small neutrino masses 
in supersymmetric models.

\end{document}